\documentclass[aps,prd,twocolumn,groupedaddress,showpacs,eqsecnum]{revtex4-1}

\usepackage{amsmath}
\usepackage{amssymb}
\usepackage{cases}
\usepackage{float}
\usepackage{mathtools}
\usepackage{keystroke}
\usepackage{rotating}
\usepackage{setspace}
\DeclareMathOperator{\sech}{sech}
\DeclareMathOperator{\cosec}{cosec}
\DeclareMathOperator{\cosech}{cosech}
\DeclareMathOperator{\csgn}{csgn}
\DeclareMathOperator{\Real}{Re}
\DeclareMathOperator{\Imag}{Im}
\newcommand{\emc}{\boldsymbol{\gamma}}
\newcommand{\abs}[1]{\left|#1\right|}
\newcommand{\la}{\left\langle}
\newcommand{\ra}{\right\rangle}
\newcommand{\ket}[1]{\left|#1\ra}
\newcommand{\bra}[1]{\la#1\right|}
\newcommand{\vab}{\bra{0}}
\newcommand{\vak}{\ket{0}}
\newcommand{\hf}[4]{F\left[#1,#2;#3;#4\right]}

\newcommand{\gf}[1]{\Gamma\left(#1\right)}
\newcommand{\psf}[1]{\psi\left(#1\right)}
\newcommand{\poch}[2]{\left(#1\right)_{#2}^{}}
\renewcommand{\Box}{\raisebox{-0.095mm}{\text{\fontsize{4mm}{1em}\selectfont{$\square$}\normalsize}}}
\newcommand{\isdef}{\mathrel{\mathop:}=}
\newcommand{\Gf}{G^{}_{\text{F}}}
\newcommand{\Gfr}{G^{}_{\text{F, reg}}}
\newcommand{\Gfs}{G^{}_{\text{F, sing}}}
\newcommand{\GfC}{G^{}_{\text{F}(C)}}
\newcommand{\GfD}{G^{}_{\text{F}(D)}}

\newcommand{\Gfh}{G^{}_{\text{H}}}
\newcommand{\Gfhr}{G^{}_{\text{H, reg}}}
\newcommand{\Gfhs}{G^{}_{\text{H, sing}}}

\newcommand{\Rmn}{R^{}_{\mu\nu}}
\newcommand{\tno}[1]{{\theta^{}_{#1}}}
\newcommand{\psr}{\langle\Phi_{}^{2}\rangle_{\text{ren}}}
\newcommand{\xx}{(x,x')}
\newcommand{\Tmn}{T^{}_{\mu\nu}}
\newcommand{\Tmnr}{\langle  T^{}_{\mu\nu}\rangle^{}_{\text{ren}}}
\newcommand{\Tmnrn}{\langle  T^{}_{\mu\nu}\rangle_{\text{ren}}}
\newcommand{\lble}[1]{\label{e:#1}}
\newcommand{\eqr}[1]{\eqref{e:#1}}
\newcommand{\lblf}[1]{\label{f:#1}}
\newcommand{\be}{\begin{equation}}
\newcommand{\ee}{\end{equation}}
\newcommand{\nno}{\nonumber}
\newcommand{\hs}{\hspace{0.25mm}}
\usepackage{color}
\usepackage{soul}

\begin{document}

\title{Hadamard renormalized scalar field theory on anti-de Sitter space-time}
\author{Carl Kent}
\email[]{c.kent@sheffield.ac.uk}
\author{Elizabeth Winstanley}
\email[]{e.winstanley@sheffield.ac.uk}
\affiliation{Consortium for Fundamental Physics,
School of Mathematics and Statistics, The University of Sheffield, Sheffield S3 7RH, United Kingdom}
\date{\today}

\begin{abstract}
We consider a real massive free quantum scalar field with arbitrary curvature coupling on $n$-dimensional anti-de Sitter space-time.
We use Hadamard renormalization to find the vacuum expectation values of the quadratic field fluctuations and the
stress-energy tensor, presenting explicit results for $n=2$ to $n=11$ inclusive.
\end{abstract}

\pacs{04.62+v}

\maketitle

\section{Introduction}
Quantum field theory (QFT) on curved space-time is a semi-classical approach to quantum gravity, where the $n$-dimensional space-time
geometry is fixed and classical, and a quantum field propagates on this background.
This approach is encapsulated in the semi-classical Einstein equations,
\be \lble{SEEs}
G^{}_{\mu\nu}+\Lambda g^{}_{\mu\nu}=8\pi\bra{\psi}\Tmn\ket{\psi}^{}_{\text{ren}},
\ee
where $\bra{\psi}\Tmn\ket{\psi}^{}_{\text{ren}}$ is the renormalized stress-energy tensor (RSET),
associated with the state $\ket{\psi}$ of a quantum field, and where we are using units in which
\be \lble{us}
c=G=\hbar=1.
\ee

The RSET is the central object of interest in QFT on curved space-time
as it is responsible for the back-reaction of the quantum field on the space-time geometry.
Since $\bra{\psi}\Tmn\ket{\psi}^{}_{\text{ren}}$ depends on the state as well as the geometry,
understanding the back-reaction requires a methodology for computing
$\bra{\psi}\Tmn\ket{\psi}^{}_{\text{ren}}$ in which the state-dependence is manifest.

The stress-energy tensor involves products of quantum operators at the same space-time point.
Hence its expectation value is
divergent and requires renormalization. Renormalization in curved space-time is challenging.
Calculations of explicit expressions for $\bra{\psi}\Tmn\ket{\psi}^{}_{\text{ren}}$
have been attempted mostly in $n=4$ space-time dimensions and appear to be tractable only for highly symmetric space-times.
A particular emphasis in the literature to date has been computations on black hole space-times.
For $n=4$, the RSET has been computed for various quantum fields on static, spherically symmetric black hole space-times~\cite{Els, How, HowCan, JenOtt, AHS1, AHS2, JenMcLOtt}, while for $n=3$ it is also possible to find the RSET on the rotating Ba\~{n}ados-Teitelboim-Zanelli (BTZ) black hole \cite{Steif, LifOrt}.
For black holes in more than four space-time dimensions, the literature on the computation of renormalized expectation values is rather limited.
There is work on computing the RSET using the Schwinger-de Witt representation of the Feynman propagator when $n$ is even \cite{Thompson:2008bk}.
In addition, the renormalized expectation value of the quadratic field fluctuations (RQFF)
has been computed on the horizon of an asymptotically flat, spherically
symmetric, three and five-dimensional black hole in \cite{MakShi} and \cite{Frolov:1989rv}, respectively, and on the five-dimensional analogue of the BTZ black hole \cite{Shiraishi:1993ti}.

The subject of QFT on anti-de Sitter space-time, $adS$, has held a particular interest for some time: initially, since $adS$ arises as a natural solution of a certain class of supergravity theories (see, for example, \cite{BDFG}); and more recently in the context of the $adS$/CFT (conformal field theory) correspondence \cite{adscft,Witten:1998qj,Gubser:1998bc,Aharony:1999ti}.
The fact that $adS$ is not globally hyperbolic \cite{H&E} has important consequences for QFT on this space-time.
Boundary conditions must be imposed at infinity in order to define a consistent QFT \cite{AS&I}.
These conditions were first explored in \cite{AS&I} for a quantum scalar field when $n=4$, leading to expressions for the scalar field modes and the Feynman Green's function in this case.
For general $n$, the scalar field modes have been obtained in \cite{B&L, Cota1} and the Feynman Green's function has been calculated by summing over these modes \cite{B&L}.
Two-point functions for scalar, spinor and vector fields have also been derived for general $n$
\cite{A&J, A&J2} by directly solving the field equations, exploiting the underlying symmetries of $adS^{}_{}$.
Vacuum expectation values for a massive scalar field with general coupling to the Ricci scalar curvature have been computed using zeta-function regularization, for $n=4$ \cite{C&H} and $n\geq2$ \cite{Cal}.  In addition, the one-loop effective potential for $n=4$ has been studied \cite{Campo}.

Hadamard renormalization (HR) was introduced in the development of Wald's axiomatic approach \cite{Wald5,Wald2}.  It extends the method of covariant geodesic point separation \cite{DeWitt, SMC} and is based on the requirement that the short-distance singularity structure of the Green's function solutions of the field equation have the form of Hadamard's `elementary solution' \cite{Had}.  The Hadamard form is a powerful device in the context of renormalization not least since its structure is given by a rigorous theorem, but moreover it clearly exhibits the influences of the field state.
Rigorous approaches to QFT in curved space rely on the notion of a Hadamard state, whose point-split two-point function has the Hadamard form.

D\'{e}canini and Folacci \cite{D&F08} have implemented HR in an elegant framework for a free scalar field coupled to a general $n$-dimensional space-time. Ref.~\cite{D&F08} serves as a practical approach to higher-dimensional renormalization in specific space-times. Even so, applications of HR to RSET computations in the literature mostly remain restricted to $n=4$ (see, for example, \cite{T&O1, B&O, T&O2, T&O3}).
In this paper our focus is on applying HR to RSET computations, but HR has wider applications, for example to studies of the
self-force~\cite{OW, CDOW} and stress-energy tensor correlators~\cite{Frob}.

In this paper, we apply HR to expectation values of operators for the \emph{global} vacuum state $\vak$ of a real massive free quantum scalar field $\Phi$ coupled to $n$-dimensional anti-de Sitter space-time, $adS^{}_{n}$. Maximally symmetric spaces such as $adS^{}_{n}$ are the simplest non-trivial space-times on which to compute the RSET.
Working on $adS^{}_{n}$ is therefore a good preliminary step in the development of practical HR methodology in more than four space-time dimensions.
Ultimately an implementation of HR for higher-dimensional black hole space-times would be of great interest.  In four space-time dimensions, a computation of the RQFF for a massless, conformally coupled scalar field on an asymptotically anti-de Sitter black hole \cite{Flachi:2008sr} has shown that the expectation value far from the black hole approaches the (non-zero) $adS_{4}$ expectation value.
Knowledge of the vacuum expectation value of the RSET for a scalar field with arbitrary mass and curvature coupling on $adS^{}_{n}$ is therefore useful for extensions of the work of \cite{Flachi:2008sr} to more general field mass and curvature coupling as well as future computations on higher-dimensional, asymptotically anti-de Sitter black holes.

The outline of this paper is as follows. We begin, in Sec.~\ref{prelim}, with a review of the geometrical foundations required for QFT on $adS^{}_{n}$ \cite{AS&I}, as well as the basic tools necessary for covariant geodesic point separation \cite{DeWitt, SMC}. Sec.~\ref{fp} is concerned with the Feynman Green's function $\Gf\xx$ for a quantum scalar field on $adS^{}_{n}$, which is constructed in Sec.~\ref{fgf}. In Sec.~\ref{hf}, we set out the Hadamard form of the Feynman Green's function, $\Gfh\xx$, which enables us, in Sec.~\ref{fpd}, to fix an overall normalization constant in $\Gf\xx$.
In order to perform HR, in Sec.~\ref{ssgf} and \ref{ssgh}, we expand both $\Gf$ and $\Gfh$ respectively as formal series up to second order in the point separation.  Based on this regularization, in Sec.~\ref{hr} we then detail the procedure used to obtain expressions for the RQFF, $\psr$, and the RSET, $\Tmnr$, with respect to $\vak$.
Analytic results for $\psr$ and $\Tmnr$ for $n=2$ to $n=11$ inclusive are presented in Sec.~\ref{res} and our conclusions are in Sec.~\ref{conc}.
In particular, we compare our results using HR with those obtained from zeta-function regularization \cite{Cal}.
Throughout this paper, we adopt the Lorentzian metric signature convention $(-,+,+,\ldots,+)$, and the unit system \eqr{us}.
An appendix lists properties of the gamma, psi and hypergeometric functions which are used in our calculations.
\vfill
\section{Preliminaries\label{prelim}}
\subsection{Scalar field theory on $adS^{}_{n}$\label{sft}}
$AdS^{}_{n}$ may be realized as the embedding of a single-sheeted $n$-dimensional hyperboloid in an {${(n+1)}$-}dimensional Euclidean space,
\be \lble{ECoords}
\mathbb{E}^{}_{n+1} = \{\zeta^{\mu}_{},\quad\mu = 0,1,2,\dots,n\},
\ee
endowed with the metric
\be \lble{EMet}
g^{\mathbb{E}}_{\mu \nu} = \text{diag}(-1,1,1,\dots,1,-1).
\ee
Intrinsic coordinates are then given by the constraint
\be \lble{Constraint}
\{x^{\mu}_{}\} = \{\zeta^{\mu}_{},\quad g^{\mathbb{E}}_{\mu \nu}\zeta^{\mu}_{}\zeta^{\nu}_{}=-a_{}^{2}\},\quad a>0,
\ee
where $ i a$ is the radius of curvature of $adS^{}_{n}$.  A convenient dimensionless coordinate system is the set of hyperspherical coordinates,
\be
\begin{array}{rcll} \lble{param}
-\pi\;<&\!\!\tau\!&\leq\,\pi,&\qquad\tau=-\pi\text{ and }\tau=\pi\text{ identified,}\\
0\;\leq&\!\!\rho\!&<\,\tfrac{\pi}{2},&\\
0\;\leq&\!\,\tno{j}&\leq\,\pi,&\qquad j=1,2,\ldots,n-3,\\
0\;\leq&\!\!\varphi\!&<\,2\pi,&
\end{array}
\ee
parametrizing the temporal, radial, polar and azimuthal directions respectively. The coordinate system \eqr{param} covers $adS^{}_{n}$, excluding polar singularities.
In terms of the coordinates \eqr{param}, the metric on $adS^{}_{n}$ takes the form
\be
ds^{2}_{} = a^{2}_{} \left(\sec\rho\right)^{2}_{}\left[ -d\tau ^{2}_{} + d\rho ^{2}_{} + \left( \sin  \rho \right)^{2}_{} d\Sigma _{n-2}^{2} \right]
\ee
where $d\Sigma _{n-2}^{2}$ is the metric on the $(n-2)$-sphere.

The simplest action for a real massive free classical scalar field $\Phi(x)$ coupled to $g^{}_{\mu\nu}$, the metric tensor of $adS^{}_{n}$, is
\be \lble{act}
S[\Phi,g^{}_{\mu\nu}]=-\frac{1}{2}\int_{\hspace{-1mm}{\text{\raisebox{-1.25mm}{$adS^{}_{n}$}}}}\hspace{-5mm}d^{n}x\,g^{\frac{1}{2}}_{}\left(g_{}^{\mu\nu}\nabla^{}_{\mu}\Phi\nabla^{}_{\nu}\Phi+m^{2}_{\xi}\Phi^{2}\right),
\ee
where
\be \lble{detg}
g\isdef\abs{\det g^{}_{\mu\nu}},
\ee
and
\be \lble{zcm}
m_{\xi}^{2}\isdef m^{2}+\xi R,
\ee
is an effective mass-squared term with $m$, the mass of the field quanta, and the constant $\xi$, the coupling between $\Phi$  and $R$, the Ricci scalar curvature. The scalar field equation of motion
\be \lble{sfwe}
\left(\Box-m_{\xi}^{2}\right)\Phi=0,
\ee
(where $\Box $ is the $n$-dimensional curved-space Laplacian) is obtained by varying $S$ \eqr{act} with respect to $\Phi$.

The quantization of \eqr{act} requires a globally hyperbolic background.  However, since $\tau=-\pi$ and $\tau=\pi$ are identified, $adS^{}_{n}$ admits closed timelike curves.  Although the covering space $C\hspace{-0.2mm}adS^{}_{n}$ (where $-\infty<\tau<+\infty$) does not have closed timelike curves, non-global hyperbolicity persists since spatial infinity $\rho=\tfrac{\pi}{2}$ remains timelike and information may be lost or gained at infinity~\cite{AS&I}.  To ensure a well-defined QFT, we adopt \emph{reflective} boundary conditions \cite{AS&I}, whereby modes of $\Phi$ vanish at $\rho=\frac{\pi}{2}$.  Such modes are known as \emph{regular} modes \cite{B&F} and exist provided
\be \lble{eta}
\eta\isdef\sqrt{m_{\xi}^{2}a_{}^{2}+\frac{(n-1)_{}^{2}}{4}}\geq0.
\ee
For the rest of this paper we shall assume that \eqr{eta} holds.
\subsection{Bitensors on $adS^{}_{n}$\label{bits}}
A \emph{bitensor} $B\xx$ is a quantity which transforms as the product of two tensors, evaluated at $x$ and $x'$ respectively \cite{SMC}.
Similarly, a \emph{biscalar} $b\xx$ transforms as the product of two scalars, evaluated at $x$ and $x'$ respectively. Covariant derivatives of $b\xx$ with respect to $x$ or $x'$,
\be \lble{nablas}
b^{}_{;\hs\mu}\isdef{\nabla^{}_{\mu}}b\xx,\qquad b^{}_{;\hs\mu'}\isdef{\nabla^{}_{\mu'}}b\xx,
\ee
are specified by an unprimed or primed index respectively.

Given two points $x,x'$ sharing a geodesic, $s\xx$ is the proper distance between them and is pure imaginary for timelike intervals.  The related invariant,
\be
\sigma\xx\isdef\frac{1}{2}\left[s\xx\right]^{2}_{},
\ee
is known as the \emph{geodetic interval} \cite{DToGF}, or \emph{world-function} \cite{Synge}, and obeys the PDE
\be
\sigma=\frac{1}{2}\sigma^{;\hs\mu}_{}\sigma_{;\hs\mu}^{}.
\ee
In this paper we are considering the maximally symmetric space-time $adS^{}_{n}$. A biscalar
$b\xx $ is a function of the proper distance $s\xx $ (that is, $b\xx =b(s)$)
if it is invariant under the action of all isometries of a maximally symmetric space-time \cite{A&J}.
This greatly simplifies our calculations.

The \emph{bivector} ${g_{\mu}^{}}_{}^{\nu'}\xx$ enables parallel transport of vectors from $x$ to $x'$ along a geodesic \cite{SMC}, such that the unit tangent vectors $s_{;\hs\mu}^{}$ and $s_{;\hs\mu'}^{}$ (at $x$ and $x'$ respectively) are related by,
\be
{g_{\mu}^{}}_{}^{\nu'}s_{;\hs\nu'}^{}=-s_{;\hs\mu}^{} .
\ee
The bivector of parallel transport satisfies the boundary condition
\be
\lim_{x'\to x}g^{}_{\mu\nu'}=g^{}_{\mu\nu}.
\ee

The Van-Vleck Morette determinant \cite{Morette, VV, Visser},
\be
\Delta\xx\isdef-\left[g(x)\right]^{-\frac{1}{2}}_{}\det\left[-\sigma\xx^{}_{;\hs\mu\nu'}\right]\left[g(x')\right]^{-\frac{1}{2}}_{},
\ee
contributes to the description of the rate of geodesic convergence \cite{DToGF}, and satisfies the PDE \cite{D&F08}
\be \lble{VVDpde}
\Box_{x}^{}\sigma=n-2\Delta^{-\frac{1}{2}}_{}{\Delta^{\frac{1}{2}}_{}}^{}_{;\hs\mu}\sigma^{;\hs\mu}_{},
\ee
where $\Box _{x}$ is the $n$-dimensional curved-space Laplacian with respect to the unprimed coordinates,
with the boundary condition
\be \lble{VVDBC}
\lim_{x'\to x}\Delta\xx=1.
\ee
The relationship
\be
\Box s=\frac{n-1}{a}\coth\left(\frac{s}{a}\right),
\ee
established in \cite{A&J} allows the explicit form of the Van-Vleck determinant to be found on $adS^{}_{n}$, namely
\be \lble{VVDAdSn}
\Delta=\left[\frac{s}{a}\cosech\left(\frac{s}{a}\right)\right]_{}^{n-1}.
\ee
\section{Scalar field Green's function\label{fp}}
\subsection{Feynman Green's function\label{fgf}}
In order to compute the RQFF and RSET, we first need to find the Feynman Green's function $\Gf\xx $ for the scalar field $\Phi $.
The Feynman Green's function is defined by
\be \lble{fp}
\Gf\xx\isdef  i\vab\mathcal{T}\left(\hs\Phi(x)\Phi(x')\right)\vak,
\ee
where
\begin{numcases}{\mathcal{T}\left(\hs\Phi(x)\Phi(x')\right)\isdef}
\Phi(x)\Phi(x'),&$t<t'$,\nno\\
\Phi(x')\Phi(x),&$t>t'$,\lble{TO}
\end{numcases}
denotes the time-ordering of the application of the field operators.
The Feynman Green's function satisfies the inhomogeneous Klein-Gordon equation
\be \lble{ISFWE1.5}
\left(\Box^{}_{x}-m_{\xi}^{2}\right)\Gf\xx=g^{-\frac{1}{2}}_{}\delta_{}^{n}(x-x').
\ee
Integral representations of $\Gf$ can be found using Schwinger's proper time method \cite{SPT}, generalized to curved space-times (see, for example,
\cite{DeWitt,DToGF,C&R, SMC}).  The time-ordered nature of $\Gf$ is guaranteed in such schemes by requiring that it be the limiting case of a function that is analytic in the upper complex half-plane of $s$.  Accordingly, in what follows, it is tacitly implied that
\be \lble{ie}
\Gf\xx\isdef\Gf(s\xx+ i0^{+}_{}),\quad0^{+}_{}\isdef\lim_{\epsilon\to0}\epsilon,\quad\epsilon>0.
\ee

Explicit expressions for $\Gf$ on $adS^{}_{n}$ can be obtained using one of two methods, both of which are greatly simplified by virtue of the underlying space-time symmetries.  Burgess and L\"{u}tken \cite{B&L} evaluate \eqr{fp} as a sum over field modes.
However, the maximal symmetry of $adS^{}_{n}$ means that we are in the special situation of being able to find a closed-form expression for the Feynman Green's function $\Gf\xx$ without employing a sum over modes.
 Allen and Jacobson \cite{A&J} show that $\Gf$ is invariant under the action of all isometries of a maximally symmetric space-time and express \eqr{ISFWE1.5} as the ODE
\be \lble{AJRE}
\left[\frac{d^{2}_{}}{ds^{2}_{}}+\frac{n-1}{a}\coth\left(\frac{s}{a}\right)\frac{d}{ds}-m_{\xi}^{2}\right]G^{}_{\text{F}}=0,
\ee
for  $s\neq 0$.  The ODE \eqr{AJRE} is then identified with the hypergeometric equation \eqr{hde}
\be \lble{AJHDE}
\left[z(1-z)\frac{d^{2}_{}}{dz^{2}_{}}-n\left(z-\frac{1}{2}\right)\frac{d}{dz}+m^{2}_{\xi}a_{}^{2}\right]G^{}_{\text{F}}=0,
\ee
for $z\neq 0$ by means of the change of variable
\be \lble{AJC}
z\isdef-\left[\sinh\left(\frac{s}{2a}\right)\right]_{}^{2}.
\ee
Following \cite{Campo} we write the solution of \eqr{AJRE} in the form (using \S9.153.7 of \cite{G&R})
\be \lble{CamGS}
G^{}_{\text{F}}=G^{}_{\text{F}(C)}+G^{}_{\text{F}(D)},
\ee
which is valid for $\abs{z}<1$, for all $n\ge 2$, where
\begin{subequations}
\begin{align}
\lble{GfC}G^{}_{\text{F}(C)}\isdef\;&C\hf{\alpha}{\beta}{\gamma}{z},\\
\lble{GfD}G^{}_{\text{F}(D)}\isdef\;&D\hf{\alpha}{\beta}{\gamma}{1-z},
\end{align}
\end{subequations}
with $C$ and $D$ coefficients to be determined, and where
\be
\alpha\isdef\frac{n-1}{2}+\eta,\qquad\beta\isdef\frac{n-1}{2}-\eta,\qquad\gamma\isdef\frac{n}{2},
\ee
with $\eta $ given by \eqr{eta}.
In view of (\ref{e:hdef}--\ref{e:AS15.1.1}), as $s\to0$,
\be
\lim^{}_{s\to 0}\GfC=C,\lble{G01}
\ee
and $\GfD$ diverges.
It follows that to isolate the short-distance singularity structure of $\Gf$, it is necessary to determine the constant $D$.

In preparation for this exercise in Sec.~\ref{fpd}, following \cite{Campo}, we apply the linear transformation \eqr{CamT1} to both \eqr{GfC} and \eqr{GfD} and then \eqr{hlt2} to the latter result, yielding the expression
\be \lble{CFSFP1}
\Gf=\lambda^{+}_{}(-z)_{}^{-\frac{n-1}{2}-\eta}F^{}_{+}(z)+\lambda^{-}_{}(-z)_{}^{-\frac{n-1}{2}+\eta}F^{}_{-}(z),
\ee
for $\abs{z}<1$, where we have defined the constants
\be
\lambda^{\pm}_{}\isdef(C+(-1)_{}^{\frac{n-1}{2}\pm\eta}D)\frac{\gf{\frac{n}{2}}\gf{\pm2\eta}}{\gf{\frac{n-1}{2}\pm\eta}\gf{\frac{1}{2}\pm\eta}},
\ee
and
\be
F^{}_{\pm}(z)\isdef\hf{\tfrac{n-1}{2}\pm\eta}{\tfrac{1}{2}\pm\eta}{1\pm2\eta}{z_{}^{-1}}.
\ee

Following \cite{Campo}, for $\Gf$ to remain finite as $s\to\infty$, it must be the case that
\be \lble{CDBC}
C=(-1)_{}^{\frac{n+1}{2}-\eta}D.
\ee
Therefore
\be \lble{Gfcampon}
\Gf=\lambda^{}_{\infty} D(-z)_{}^{-\frac{n-1}{2}-\eta}F^{}_{+}(z),
\ee
where
\be
\lambda^{}_{\infty}\isdef2 i_{}^n\sin\left(\pi\eta\right)\frac{\gf{\frac{n}{2}}\gf{-2\eta}}{\gf{\frac{n-1}{2}-\eta}\gf{\frac{1}{2}-\eta}}.
\ee
\smallskip

\subsection{Hadamard form\label{hf}}
The Hadamard form of $\Gf$ on $adS^{}_{n}$ is given by (recalling \eqr{ie})
\be\lble{Hadx}
\Gfh\xx= i\kappa\left[U\xx\sigma_{}^{1-\frac{n}{2}}+V\xx \ln\bar{\sigma}+W\xx \right],
\ee
with
\begin{subnumcases}{\kappa\isdef}
\displaystyle\frac{1}{4\pi},\hphantom{\hspace{-0.5mm}}\hphantom{\displaystyle\sum_{l=0}^{\frac{n}{2}-2}U^{}_{l}(\sigma)\sigma_{}^{l}}&$n=2$,\\
\displaystyle\frac{\gf{\frac{n}{2}-1}}{2(2\pi)_{}^{\frac{n}{2}}},&$n>2$,
\end{subnumcases}
and where $U\xx$, $V\xx$ and $W\xx$ are biscalars which are regular as $x'\to x$.
In \eqr{Hadx} we have defined the dimensionless quantity
\be \lble{bar}
\bar{\sigma}\isdef M^{2}_{}\sigma ,
\ee
which depends on an undetermined \emph{renormalization mass scale} $M$ \cite{D&F08}.

The biscalars $U\xx $ and $V\xx $ are purely geometric, depending on the space-time geometry but not on the state of the quantum field.
Since we are working on the maximally symmetric $adS^{}_{n}$ space-time, $U\xx$ and $V\xx$ are therefore functions of the geodetic interval $\sigma \xx $ (or, equivalently, $s\xx $).
The conventional \emph{Ansatze} for the functions $U$ and $V$ are the series expansions
\begin{subnumcases}{\label{e:Usig}U(\sigma)=}
0,&\quad$n=2,$\lble{Usiga}\\
\displaystyle\sum_{l=0}^{+\infty}U^{}_{l}(\sigma)\sigma_{}^{l},&\quad$n\text{ odd},$\lble{Usigb}\\
\displaystyle\sum_{l=0}^{\frac{n}{2}-2}U^{}_{l}(\sigma)\sigma_{}^{l},&\quad$n>2,\;n\text{ even,}$\quad\lble{Usigc}
\end{subnumcases}
and
\begin{subnumcases}{\label{e:Vsig}V(\sigma)=}
0,&\quad$n\text{ odd},$\lble{Vsiga}\\
\mathrlap{\displaystyle\sum_{l=0}^{+\infty}V^{}_{l}(\sigma)\sigma_{}^{l},}{\phantom{\displaystyle\sum_{l=0}^{\frac{n}{2}-2}U^{}_{l}(\sigma)\sigma_{}^{l},}}&\quad$\mathrlap{n\text{ even.}}{\phantom{n>2,\;n\text{ even.}}}$\quad\lble{Vsigb}
\end{subnumcases}

The following recursion relations determine the coefficients $U^{}_{l}$ and $V^{}_{l}$
\begin{subequations}
\label{e:Hcoeffs}
\begin{align}
\frac{\Box^{}_{x}-m_{\xi}^{2}}{2l+4-n}U^{}_{l}=\;&U^{}_{l+1}\Delta^{-\frac{1}{2}}_{}\Delta_{}^{\frac{1}{2}}{}^{}_{;\hs\mu}\hs\sigma^{;\hs\mu}_{}\nno\\
&\qquad-(l+1)U^{}_{l+1}-U^{}_{l+1;\hs\mu}\sigma_{}^{;\hs\mu},\lble{Hcoeffsa}\\
\frac{\Box^{}_{x}-m_{\xi}^{2}}{l+1}V^{}_{l}=\;&2V^{}_{l+1}\Delta_{}^{-\frac{1}{2}}\Delta_{}^{\frac{1}{2}}{}^{}_{;\hs\mu}\hs\sigma_{}^{;\hs\mu}\nno\\
&\qquad-(2l+n)V^{}_{l+1}-2V^{}_{l+1;\hs\mu}\sigma^{;\hs\mu}_{},\lble{Hcoeffsb}
\end{align}
\end{subequations}
\noindent
with the associated boundary conditions for $U^{}_{0}$ and $V^{}_{0}$:
\begin{equation}\lble{HUBC}
U^{}_{0}=\Delta_{}^{\frac{1}{2}},
\end{equation}
and
\begin{widetext}
\begin{subnumcases}{V^{}_{0}=\label{e:HVBC}}
-\Delta_{}^{\frac{1}{2}},&$n=2$,\lble{HVBCa}\\
\frac{1}{(n-2)}\left[2V^{}_{0}\Delta_{}^{-\frac{1}{2}}\Delta_{}^{\frac{1}{2}}{}^{}_{;\hs\mu}\hs\sigma_{}^{;\hs\mu}-2V^{}_{0;\hs\mu}\sigma_{}^{;\hs\mu}-\left(\Box^{}_{x}-m_{\xi}^{2}\right)U^{}_{\frac{n-4}{2}}\right],\quad&$n>2$.\quad\lble{HVBCb}
\end{subnumcases}
\end{widetext}
Using these relations, the $U^{}_{l}$ and $V^{}_{l}$  (and hence $U(\sigma )$ and $V(\sigma )$) can be determined by integrating along the unique geodesic separating the events $x$ and $x'$ (assuming that $x$ and $x'$ are sufficiently close that they are connected by a unique geodesic).  However, the biscalar $W\xx$ is not uniquely defined, a feature that can be attributed to the state-dependence of $\Gfh$.  It follows that the terms of \eqr{Hadx} involving the functions $U$ and $V$ (i.e. those terms containing the short-distance singularities), are the purely geometrical parts of $\Gfh$, and the term of \eqr{Hadx} involving the biscalar $W\xx$ additionally incorporates the state-dependence of $\Gfh$.
Even on the maximally symmetric $adS^{}_{n}$ space-time, for a general quantum state the biscalar $W\xx $ is not simply a function of the geodetic interval $\sigma \xx$.
However, if we are considering a quantum state which shares the maximal symmetry of the underlying $adS^{}_{n}$ space-time, then $W\xx $ will depend on $x$ and $x'$ only through $\sigma \xx$, so we can write $W\xx = W(\sigma )$.
In particular, this is the case for the global $adS^{}_{n}$ vacuum which is our focus in this paper.
\subsection{Determination of the coefficient $D$\label{fpd}}
We now complete our determination of the Feynman Green's function $\Gf$ by finding the coefficient $D$ in \eqr{Gfcampon}.
This coefficient is fixed by directly comparing the leading-order divergence of $\Gf$ with that of $\Gfh$. To extract the leading-order divergence of $\Gf$, we follow \cite{A&J}.
Applying \eqr{hlt2} to the hypergeometric function in \eqr{Gfcampon} gives
\be \lble{GcampT1}
F_{+}(z)=\left(1-z_{}^{-1}\right)_{}^{-\frac{n-1}{2}-\eta}F(\tilde{z}),
\ee
where
\be \lble{Fztilde}
F(\tilde{z})=\hf{\tfrac{n-1}{2}+\eta}{\tfrac{1}{2}+\eta}{2\eta+1}{\tilde{z}},
\ee
with
\be
\tilde{z}\isdef\frac{1}{1-z} = \left[ \sech \left( \frac {s}{2a} \right) \right] ^{2}.
\ee
As $z\to 0$, we have ${\tilde {z}}\to 1$.
Applying \eqr{hlt1} to \eqr{Fztilde} gives
\be \lble{GcampT2}
F(\tilde{z})=\left(1-\tilde{z}\right)_{}^{1-\frac{n}{2}}\tilde{F}(\tilde{z}),
\ee
where
\be
\tilde{F}(\tilde{z})\isdef\hf{\eta-\tfrac{n-3}{2}}{\tfrac{1}{2}+\eta}{2\eta+1}{\tilde{z}},
\ee
is convergent as ${\tilde {z}}\to 1$ for all $n>2$. Combining (\ref{e:Gfcampon}, \ref{e:GcampT1}, \ref{e:GcampT2}) gives
\be \lble{GfcamponTng2}
G_{\text{F}}^{n>2}=\lambda^{}_{\infty}\tilde{z}_{}^{\frac{n-1}{2}+\eta}(1-\tilde{z})_{}^{1-\frac{n}{2}}\tilde{F}(\tilde{z}),
\ee
where all the singular behaviour in $\Gf$ as ${\tilde {z}}\to 1$ ($z\to 0$) is now in the $(1-\tilde{z})_{}^{1-\frac{n}{2}}$ factor.

When $n=2$, given \eqr{GcampT1} and applying the transformation \eqr{hlt4} to $F\left(\tilde{z}\right)$,
the Green's function \eqr{Gfcampon} has the form
\begin{align}
\lble{GfcamponTne2}
G_{\text{F}}^{n=2}=\;&\lambda^{}_{\infty} D\tilde{z}_{}^{\eta+\frac{1}{2}}\frac{\gf{2\eta+1}}{\left[\gf{\frac{1}{2}+\eta}\right]_{}^2}\nno\\
&\times\sum_{k=0}^{+\infty}\left[\frac{\poch{\frac{1}{2}+\eta}{k}}{k!}\right]_{}^2\!\left[\Psi^{n=2}-\ln(1-\tilde{z})\right]\!(1-\tilde{z})_{}^{k},
\end{align}
where $\poch{\frac{1}{2}+\eta}{k}$ is the Pochhammer symbol defined in \eqr{pdef} and where
\be \lble{Psi1n2}
\Psi^{n=2}\isdef2\left[\psf{k+1}-\psf{\tfrac{1}{2}+\eta+k}\right] ,
\ee
with $\psi$ denoting the psi function defined in \eqr{psidef}.
The expression \eqr{GfcamponTne2} exhibits a logarithmic leading-order divergence.

With the expressions (\ref{e:GfcamponTng2}--\ref{e:GfcamponTne2}) in hand, the coefficient $D$ can be determined by extracting the leading-order divergences and then directly comparing this behaviour with that of their respective Hadamard form counterparts,
\begin{subequations}
\begin{align}
\lble{Ghlon=2}G_{\text{H}}^{n=2}&\;\sim-\frac{ i}{2\pi}\ln\bar{s},&s\to0,\\
\lble{Ghlon>2}G_{\text{H}}^{n>2}&\;\sim\frac{ i}{4\pi_{}^{\frac{n}{2}}}\gf{\tfrac{n}{2}-1}s_{}^{2-n},\qquad&s\to0,
\end{align}
\end{subequations}
obtained from (\ref{e:Hadx}, \ref{e:HUBC}, \ref{e:HVBC}) with \eqr{VVDBC}, where `$\sim$' denotes `diverges as' and defining $\bar {s} \isdef M s$
where $M$ is the renormalization mass scale \eqr{bar}.

When $n=2$, as $s\to0$ the $k=0$ terms are the only contributors to the summation in \eqr{GfcamponTne2}.  Furthermore,
\be \lble{GfDne2div}
\ln \left(1- {\tilde {z}} \right) = \ln\left\{\left[\tanh\left(\frac{s}{2a}\right)\right]_{}^2\right\}\sim2\ln\bar{s},\qquad s\to0.
\ee
Making use of \eqr{gref}, the comparison of \eqr{GfcamponTne2} and \eqr{Ghlon=2} then yields
\be \lble{Dne2u}
D=\frac{i}{4}\sec(\pi\eta),
\ee
for $n=2$.

When $n>2$, the leading-order divergence of $G^{n>2}_{\text{F}}$ is that of the
$\left( 1 - {\tilde {z}} \right) ^{1-\frac {n}{2}}=\left[\tanh\left(\frac{s}{2a}\right)\right]_{}^{2-n}$ factor in \eqr{GfcamponTng2}. Thus, as $s\to0$,
\be \lble{GfDng2div}
G_{\text{F}}^{n>2}\sim \lambda^{}_{\infty} D\frac{\gf{\frac{n}{2}-1}\gf{2\eta+1}}{\gf{\frac{n-1}{2}+\eta}\gf{\frac{1}{2}+\eta}}(2a)_{}^{n-2}s_{}^{2-n}.
\ee
Making use of \eqr{gref} and \eqr{pdef}, for $p=1,2,\ldots$, the comparison of \eqr{Ghlon>2} and \eqr{GfDng2div} then yields, for $n>2$,
\be \lble{Dnou}
D=\frac{(-1)_{}^{p}}{(4\pi)_{}^{p+\frac{1}{2}}a_{}^{2p-1}\gf{p+\frac{1}{2}}}\frac{\poch{-\eta}{p}\poch{\eta}{p}}{\eta}\pi\cosec(\pi\eta),
\ee
for $n=2p+1$ and
\be \lble{Dneu}
D=\frac{i(-1)_{}^{p}}{(4\pi)_{}^{p+1}a_{}^{2p}p!}\poch{\tfrac{1}{2}-\eta}{p}\poch{\tfrac{1}{2}+\eta}{p}\pi\sec(\pi\eta),
\ee
for $n=2p+2$.

We emphasize at this stage that despite the presence of poles in \eqr{Dne2u} and \eqr{Dneu} for $\eta=\tfrac{1}{2},\tfrac{3}{2},\ldots$, and in \eqr{Dnou} for $\eta=1,2,\ldots$, the Feynman Green's function $\Gf$ remains finite for all $\eta$.
\section{Singularity structure of $G^{}_{\mathrm{F}}$\label{ssgf}}
To compute the RQFF and RSET using HR, the purely geometric, divergent parts of the Hadamard form \eqr{Hadx} (i.e.~the terms containing $U$ and $V$) are subtracted from the Feynman Green's function \eqr{CamGS} and the limit {$x'\to x$} is taken.
The first step is therefore to \emph{regularize}, by isolating all the divergent terms in both the Feynman Green's function \eqr{CamGS} and the Hadamard form \eqr{Hadx} and write them in a form which facilitates \emph{renormalization}, i.e.~the subtraction and limit process.
In this section we regularize the Feynman Green's function \eqr{CamGS} and in the next section, the corresponding calculations for the Hadamard form \eqr{Hadx} are performed.

The regularization
\be \lble{psrregsing}
\Gf=\Gfr+\Gfs,
\ee
where $\Gfs$ denotes the part of $\Gf$ containing its singular terms and $\Gfr$ denotes the remaining regular part of $\Gf$ is developed below by expanding the hypergeometric functions in \eqr{GfC} and \eqr{GfD} as series in $s$, for $n=2$ and $n>2$ with $n$ odd and with $n$ even respectively.

The computations of $\psr$ and $\Tmnr$ require an expression of $\Gf$ to zeroth and second order in $s$ as their corresponding inputs.  In what follows, we adopt a convenient notation for such expressions. For a formal series $f(s)$, we denote: $f^{\boldsymbol{0}}_{}$ and $f^{\textbf{II}}_{}$ as its respective zeroth and second order truncations; $f^{(0)}_{}$ and $f^{(2)}_{}$ as its zeroth and second order terms; and $f^{(*)}_{}$ as its singular terms.
\subsection{$n=2$\label{GF2}}
The relations (\ref{e:CDBC}, \ref{e:Dne2u}, \ref{e:gref}) and the expansion \eqr{hdef} of the hypergeometric function in \eqr{GfC} combine to give
\be \lble{GfC2po2}
G_{\text{F}(C)}^{n=2}=\frac{1}{4}\left[1- i\tan(\pi\eta)\right]\sum_{k=0}^{+\infty}\frac{\poch{\frac{1}{2}+\eta}{k}\poch{\frac{1}{2}-\eta}{k}}{\left(k!\right)_{}^2}z_{}^{k}.
\ee

Recalling (\ref{e:Dne2u}, \ref{e:gref}) and applying \eqr{hlt4} to the hypergeometric function in \eqr{GfD} yields
\be \lble{GfD2}
G^{n=2}_{\text{F}(D)}=\frac{ i}{4\pi}\sum_{k=0}^{+\infty}\frac{\poch{\frac{1}{2}+\eta}{k}\poch{\frac{1}{2}-\eta}{k}}{\left(k!\right)^2}\left(\tilde{\Psi}^{n=2}_{}-\ln z\right)z^k_{},
\ee
valid as $s\to0$, where
\be \lble{Psi2.2}
\tilde{\Psi}^{n=2}_{}\isdef2\psi(k+1)-\psi\left(\tfrac{1}{2}+\eta+k\right)-\psi\left(\tfrac{1}{2}-\eta+k\right).
\ee
The Feynman Green's function therefore has a logarithmic singularity as $s\to 0$, which can be seen explicitly from
\be \lble{lnzexp2}
\ln z=2\ln\bar{s}-2\ln2-2\ln\bar{a}- i\csgn\left( i s^{2}_{}\right)\pi+\frac{s^{2}_{}}{12a^{2}_{}}+O\!\left(s^{4}_{}\right),\hfill
\ee
as $s\to0$, where
\be
\csgn{y}\isdef
\begin{cases}
\phantom{-}1,&\Real y>0,\quad\text{or}\quad\Real y=0,\;\Imag y>0,\\
\phantom{-}0,&y=0,\\
-1,&\Real y<0,\quad\text{or}\quad\Real y=0,\;\Imag y<0,
\end{cases}
\ee
is the generalization of the signum function to complex numbers $y$ \cite{Maple}, and where we have defined the dimensionless variable $\bar{a}\isdef Ma$
where $M$ is the renormalization mass scale \eqr{bar}.

Adding \eqr{GfC2po2} and \eqr{GfD2} gives
\be \lble{Gf2se}
G_{\text{F}}^{n=2}=\frac{1}{4\pi}\sum_{k=0}^{+\infty}\frac{\poch{\frac{1}{2}+\eta}{k}\poch{\frac{1}{2}-\eta}{k}}{\left(k!\right)_{}^2}Z_{\text{log}}^{n=2},
\ee
where
\begin{align}
\lble{Z2}
Z_{\text{log}}^{n=2}\isdef\left\{\pi- i\left[2\psf{\tfrac{1}{2}+\eta}+2\emc+L+\ln z\right]\right\}z_{}^{k},
\end{align}
with the \emph{boldface} $\emc$ being the Euler-Mascheroni constant and
\be
\lble{L2}
L\isdef\sum_{l=0}^{k-1}\left(\frac{1}{\frac{1}{2}+\eta+l}+\frac{1}{\frac{1}{2}-\eta+l}\right)-2\sum_{l=1}^{k}\frac{1}{l},
\ee
having applied \eqr{psirec} and \eqr{psiref} to \eqr{Psi2.2}.

For the summation in \eqr{Gf2se}, only the $k=0$ and $k=1$ terms contribute to a second-order expansion in $s$.
From (\ref{e:Z2}, \ref{e:L2}), the contributions from $Z_{\text{log}}^{n=2}$ to these terms are
\be
Z_{\text{log}}^{n=2}|^{}_{k=0}=
\pi-i\left[2\psf{\tfrac{1}{2}+\eta}+2\emc+\ln z\right],
\ee
and
\be\lble{Z2-1}
Z_{\text{log}}^{n=2}|^{}_{k=1}=\left(Z_{\text{log}}^{n=2}|^{}_{k=0}-i\delta L\right)z,
\ee
where
\be
\lble{dL2}
\delta L\isdef L|^{}_{k=1}-L|^{}_{k=0}=L|^{}_{k=1}=\frac{\tfrac{1}{2}+2\eta^{2}_{}}{\tfrac{1}{4}-\eta^{2}_{}}.
\ee
Recalling \eqr{lnzexp2}, the second-order truncation of the $k=0$ term is
\be
Z_{\text{log}}^{\textbf{II},\,n=2}|^{}_{k=0}=-2i\left(\mathit{\Upsilon}+\ln\bar{s}+\frac{s^{2}_{}}{24a^{2}_{}}\right),
\ee
where
\be
\lble{ups}
\mathit{\Upsilon}\isdef\psf{\tfrac{1}{2}+\eta}+\emc-\ln2-\ln \bar{a}.
\ee
From \eqr{Z2-1} the second-order truncation of the $k=1$ term is then
\be
Z_{\text{log}}^{\textbf{II},\,n=2}|^{}_{k=1}=\left(Z_{\text{log}}^{\textbf{II},\,n=2}|^{}_{k=0}-i\delta L\right)\frac{s^{2}_{}}{4a^{2}_{}},
\ee

The truncation of $G_{\text{F}}^{n=2}$ to second order in $s$ is therefore
\begin{align} \lble{Gf2II}
 G_{\text{F}}^{\textbf{II},\,n=2}
=\;&-\frac{ i}{2\pi}\left\{\left(\mathit{\Upsilon}+\ln\bar{s}+\frac{s^{2}_{}}{24a^{2}_{}}\right)\right.\nno\\
&\qquad\left.\times\left[1+\left(\tfrac{1}{4}-\eta^{2}\right)\frac{s^{2}_{}}{4a^{2}_{}}\right]
+\left(\tfrac{1}{4}+\eta^{2}\right)\frac{s^{2}_{}}{4a^{2}_{}}\right\}.
\end{align}
Accordingly, the zeroth order terms of $\Gf$ are
\be \lble{Gf2s0r}
G_{\text{F}}^{(0),\,n=2}=G_{\text{F, reg}}^{\hs\boldsymbol{0},\,n=2}=-\frac{ i}{2\pi}\mathit{\Upsilon},
\ee
and the singular term of $\Gf$ is
\be \lble{Gf2s0s}
G_{\text{F}}^{(*),\,n=2}=G_{\text{F, sing}}^{\hs\boldsymbol{0},\,n=2}=-\frac{ i}{2\pi}\ln\bar{s}.
\ee
\subsection{$n>2$, $n$ even\label{GFe}}
The relations (\ref{e:CDBC}, \ref{e:Dneu}, \ref{e:gref}, \ref{e:pd1}) and the expansion \eqr{hdef} of the hypergeometric function in \eqr{GfC} combine to give
\begin{widetext}
\begin{align} \lble{GfCepo}
G_{\text{F}(C)}^{n=2p+2}=&\;\frac{\pi\left[1- i\tan(\pi\eta)\right]}{\left(4\pi\right){\!}_{}^{p+1}a_{}^{2p}}\poch{\tfrac{1}{2}+\eta}{p}\poch{\tfrac{1}{2}-\eta}{p}
\sum_{k=0}^{+\infty}\frac{\poch{p+\frac{1}{2}+\eta}{k}\poch{p+\frac{1}{2}-\eta}{k}}{k!\left(k+p\right)!}z_{}^{k}.
\end{align}

Recalling \eqr{Dneu}, applying \eqr{hlt5} to the hypergeometric function present in \eqr{GfD}, then simplifying with \eqr{gref} yields
\begin{align}
\lble{GfDe}
G^{n=2p+2}_{\text{F}(D)}
=\frac{-i}{\left(4\pi\right)_{}^{p+1}a_{}^{2p}}&\left[(-1)_{}^{p}\gf{p}z_{}^{-p}\sum_{k=0}^{p-1}\frac{\poch{\frac{1}{2}+\eta}{k}\poch{\frac{1}{2}-\eta}{k}}{k!\poch{1-p}{k}}z_{}^k
\right.\nno\\
&\qquad\left.{+\poch{\tfrac{1}{2}+\eta}{p}\poch{\tfrac{1}{2}-\eta}{p}}
\sum_{k=0}^{+\infty}\frac{\poch{p+\frac{1}{2}+\eta}{k}\poch{p+\frac{1}{2}-\eta}{k}}{k!(k+p)!}
{\left(\Psi_{}^{n=2p+2}+\ln z\right)z_{}^k}\right]\!,
\end{align}
where
\be
\lble{Psie}
\Psi_{}^{n=2p+2}\isdef\;-\psi(k+1)-\psi(k+p+1)
+\psi\left(p+\tfrac{1}{2}+\eta+k\right)+\psi\left(p+\tfrac{1}{2}-\eta+k\right).
\ee
\end{widetext}
Eq.~\eqr{GfDe} reveals poles in the formal Laurent series (FLS) of the first summation and a logarithmic divergence in the second summation.

Adding \eqr{GfCepo} and \eqr{GfDe} gives the Feynman Green's function $G_{\text{F}}^{n=2p+2}$.
To expand $G_{\text{F}}^{n=2p+2}$ to second order in $s$, it is convenient to split it into a part containing the logarithmically divergent terms and another containing the FLS:
\be
G_{\text{F}}^{n=2p+2} = G_{\text{F, log}}^{n=2p+2} + G_{\text{F, \textsc{fls}}}^{n=2p+2}
\lble{Gfn2p+2}
\ee
where
\begin{align}
\lble{Gfelog}G_{\text{F, log}}^{n=2p+2}\isdef\;\!&\frac{1}{\left(4\pi\right)_{}^{p+1}a_{}^{2p}}\poch{\tfrac{1}{2}+\eta}{p}\poch{\tfrac{1}{2}-\eta}{p}\nno\\
&\times\!\sum_{k=0}^{+\infty}\frac{\poch{p+\frac{1}{2}+\eta}{k}\!\poch{p+\frac{1}{2}-\eta}{k}}{k!\left(k+p\right)!}Z_{\text{log}}^{n=2p+2},
\end{align}
and
\be
\lble{Gfefls}G_{\text{F, \textsc{fls}}}^{n=2p+2}=\frac{1}{\left(4\pi\right)_{}^{p+1}a_{}^{2p}}Z_{\textsc{fls}}^{n=2p+2}.
\ee
In Eqs.~(\ref{e:Gfelog}--\ref{e:Gfefls}), we have introduced the quantity
\be \lble{Zesimp}
Z_{\text{log}}^{n=2p+2}\isdef\left\{\pi- i\left[2\psf{\tfrac{1}{2}+\eta}+2\emc+\widetilde{L}+\ln z\right]\right\}z_{}^{k},
\ee
where
\be \lble{Le}
\widetilde{L}\isdef\sum_{l=0}^{p+k-1}\left(\frac{1}{\frac{1}{2}+\eta+l}+\frac{1}{\frac{1}{2}-\eta+l}\right)-\sum_{l=1}^{k+p}\frac{1}{l}-\sum_{l=1}^{k}\frac{1}{l},
\ee
(having applied \eqr{psirec} and \eqr{psiref} to \eqr{Psie}), and where
\be \lble{Se}
Z_{\textsc{fls}}^{n=2p+2}\isdef i\left(-1\right)_{}^{p}\gf{p}\sum_{k=0}^{p-1}\frac{\poch{\frac{1}{2}+\eta}{k}\poch{\frac{1}{2}-\eta}{k}}{k!\poch{1-p}{k}}z_{}^{k-p}.
\ee

For the summation in \eqr{Gfelog}, only the $k=0$ and $k=1$ terms contribute to a second-order expansion in $s$.
From (\ref{e:Zesimp}, \ref{e:Le}), the $Z_{\text{log}}^{n=2p+2}$ contributions to these terms are
\be
Z_{\text{log}}^{n=2p+2}|^{}_{k=0}=
\pi-i\left[2\psf{\tfrac{1}{2}+\eta}+2\emc+\widetilde{L}|^{}_{k=0}+\ln z\right],
\ee
and
\be \lble{Ze-1}
Z_{\text{log}}^{n=2p+2}|^{}_{k=1}=\left(Z_{\text{log}}^{n=2p+2}|^{}_{k=0}-i\delta\widetilde{L}\right)z,
\ee
where
\be
\delta\widetilde{L}\isdef\widetilde{L}|^{}_{k=1}-\widetilde{L}|^{}_{k=0}=\frac{2p+1}{\left(p+\frac{1}{2}\right)^{2}_{}-\eta^{2}_{}}-\frac{p+2}{p+1}.
\ee
Recalling \eqr{lnzexp2}, the second-order truncation of the $k=0$ term is
\begin{align}
Z_{\text{log}}^{\hs\textbf{II},\,n=2p+2}|^{}_{k=0}=&-2 i\left(\mathit{\Upsilon}+\frac{1}{2}\widetilde{L}|^{}_{k=0}+\ln\bar{s}+\frac{s^{2}_{}}{12a^{2}_{}}\right).
\end{align}
From \eqr{Ze-1}, the second-order truncation of the $k=1$ term is then
\be
Z_{\text{log}}^{\hs\textbf{II},\,n=2p+2}|^{}_{k=1}=\left(Z_{\text{log}}^{\hs\textbf{II},\,n=2p+2}|^{}_{k=0}-i\delta\widetilde{L}\right)\frac{s^{2}_{}}{4a^{2}_{}}.
\ee
The truncation of the expansion of $\Gf$ to second order in $s$ is therefore:
\vspace{-1cm}
\begin{widetext}
\begin{align} \lble{GfeII}
G_{\text{F}}^{\hs\textbf{II},\,n=2p+2}=\;&\frac{1}{\left(4\pi\right)_{}^{p+1}a_{}^{2p}(p+1)!}\poch{\tfrac{1}{2}+\eta}{p}\poch{\tfrac{1}{2}-\eta}{p}\nno\\
&\times\left[Z_{\text{log}}^{\hs\textbf{II},\,n=2p+2}|^{}_{k=0}\left(p+1-\left[\eta^{2}_{}-\left(p+\tfrac{1}{2}\right)^{2}_{}\right]\frac{s^{2}_{}}{4a^{2}_{}}\right)+i\delta\widetilde{L}\left[\eta^{2}_{}-\left(p+\tfrac{1}{2}\right)^{2}_{}\right]\frac{s^{2}_{}}{4a^{2}_{}}\right]+\frac{Z_{\textsc{fls}}^{\hs\textbf{II},\,n=2p+2}}{\left(4\pi\right)_{}^{p+1}a_{}^{2p}},
\end{align}
Accordingly, recalling \eqr{lnzexp2}, the zeroth order terms of $\Gfr$ are
\begin{align}
\lble{Gfez0r}
G_{\text{F, reg}}^{(0),\,n=2p+2}=\;&\frac{- i}{4_{}^{p+\frac{1}{2}}\pi_{}^{p+1}a_{}^{2p}p!}\poch{\tfrac{1}{2}+\eta}{p}\poch{\tfrac{1}{2}-\eta}{p}
\left[\psi\left(\tfrac{1}{2}+\eta\right)+\emc+\frac{1}{2}\widetilde{L}|^{}_{k=0}-\ln2-\ln\bar{a}\right],
\end{align}
and the zeroth order truncation of the part of $\Gf$ containing its singularities is
\begin{align}
\lble{Gfez0s}
G_{\text{F, sing}}^{\hs\boldsymbol{0},\,n=2p+2}
=\frac{1}{\left(4\pi\right)_{}^{p+1}a_{}^{2p}}
\left[Z_{\textsc{fls}}^{\hs\boldsymbol{0},\,n=2p+2}|^{}_{k=0}-2i\frac{\poch{\frac{1}{2}+\eta}{p}\poch{\frac{1}{2}-\eta}{p}}{p!}\ln\bar{s}\right].
\end{align}
\end{widetext}

Recalling \eqr{Se} and in particular that the $k$\textsuperscript{th} summands for $k=0,1,\ldots,p-1$, are each respectively a FLS of leading order $s^{-2p}_{},s^{2-2p}_{},\ldots,s^{-2}_{}$, it follows that as $s\to0$, the sum $Z^{\hs\boldsymbol{0},\,n=2p+2}_{\textsc{fls}}$ also contains non-vanishing finite terms.
These non-zero finite terms contribute to the singular part of the Feynman Green's function, $\Gfs$.
For the purposes of renormalization in Sec.~\ref{hr}, it is therefore useful to make a further decomposition of \eqr{Gfez0s}:
\be
G_{\text{F, sing}}^{\hs\boldsymbol{0}}=G_{\text{F, sing}}^{(*)}+G_{\text{F, sing}}^{(0)},
\ee
where $G_{\text{F, sing}}^{(*)}$ contains all the singular terms and $G_{\text{F, sing}}^{(0)}$ is finite and non-zero as $s\to 0$.

\begin{widetext}
\subsection{$n>2$, $n$ odd\label{GFo}}
The relations (\ref{e:CDBC}, \ref{e:Dnou}, \ref{e:gref}, \ref{e:pd1}) and the hypergeometric function in \eqr{GfC} combine to give
\be
\lble{GfCopo}
G_{\text{F}(C)}^{n=2p+1}=\frac{1}{\left(4\pi\right){\!}_{}^{p+\frac{1}{2}}a_{}^{2p-1}\gf{p+\frac{1}{2}}}\frac{\poch{\eta}{p}\poch{-\eta}{p}}{\eta}
{\left[i-\cot(\pi\eta)\right]\pi\hf{p+\eta}{p-\eta}{p+\tfrac{1}{2}}{z}}.
\ee
Recalling \eqr{Dnou}, applying \eqr{hlt3} to the hypergeometric function in \eqr{GfD}, and then applying the results (\ref{e:gref}, \ref{e:pd1})  yields
\be
\lble{GfDoposimp}
G_{\text{F}(D)}^{n=2p+1}
=\frac{1}{(4\pi)_{}^{p+\tfrac{1}{2}}a_{}^{2p-1}}
\left\{\frac{1}{\gf{p+\frac{1}{2}}}\frac{\poch{\eta}{p}\poch{-\eta}{p}}{\eta}\pi\cot(\pi\eta)
{\hf{p+\eta}{p-\eta}{p+\tfrac{1}{2}}{z}}+Z_{\textsc{fls}}^{n=2p+1}\right\},
\ee
where
\be
\lble{So}
Z_{\textsc{fls}}^{n=2p+1}\isdef-\left(-1\right)^{p}_{}\gf{p-\tfrac{1}{2}}
\sum_{k=0}^{+\infty}\frac{\poch{\frac{1}{2}+\eta}{k}\poch{\frac{1}{2}-\eta}{k}}{\poch{\frac{3}{2}-p}{k}}\frac{z_{}^{k+\frac{1}{2}-p}}{k!}.
\ee
Adding \eqr{GfCopo} and \eqr{GfDoposimp} gives
\be
\lble{Gfose}
G_{\text{F}}^{n=2p+1}=\frac{1}{(4\pi)_{}^{p+\frac{1}{2}}a_{}^{2p-1}}
\left\{\frac{ i\pi}{\gf{p+\frac{1}{2}}}\frac{\poch{\eta}{p}\poch{-\eta}{p}}{\eta}
{\hf{p+\eta}{p-\eta}{p+\tfrac{1}{2}}{z}}+Z_{\textsc{fls}}^{n=2p+1}\right\},
\ee
Recalling \eqr{hdef}, the truncation of the expansion of $\Gf$ to second order in $s$ is given by
\be
\lble{GfoII}
G_{\text{F}}^{\hs\textbf{II},\,n=2p+1}=\frac{1}{(4\pi)_{}^{p+\frac{1}{2}}a_{}^{2p-1}}
\left\{\frac{ i\pi}{\gf{p+\frac{1}{2}}}\frac{\poch{\eta}{p}\poch{-\eta}{p}}{\eta}\left[1-\frac{(p+\eta)(p-\eta)}{p+\frac{1}{2}}\frac{s^{2}_{}}{4a^{2}_{}}\right]
{+Z_{\textsc{fls}}^{\hs\textbf{II},\,n=2p+1}}\right\}\!.
\ee
\end{widetext}
Accordingly, the zeroth order terms of $\Gf$ are
\be \lble{Gfoz0r}
G_{\text{F}}^{(0),\,n=2p+1}=\frac{i\pi}{(4\pi)_{}^{p+\frac{1}{2}}a_{}^{2p-1}\gf{p+\frac{1}{2}}}\frac{\poch{\eta}{p}\poch{-\eta}{p}}{\eta},
\ee
and the singular terms of $\Gf$ are
\be \lble{Gfoz0s}
G_{\text{F}}^{(*),\,n=2p+1}=G_{\text{F, sing}}^{\hs\boldsymbol{0},\,n=2p+1}=\frac{1}{(4\pi)_{}^{p+\frac{1}{2}}a_{}^{2p-1}}Z_{\textsc{fls}}^{\hs\boldsymbol{0},\,n=2p+1}.
\ee
Recalling \eqr{So} and in particular that the $k$\textsuperscript{th} summands for $k=0,1,\ldots,p-1$, are each respectively a FLS of leading order $s^{1-2p}_{},s^{3-2p}_{},\ldots,s^{-1}_{}$, it follows that as {$s\to0$}, the sum $Z^{\hs\boldsymbol{0},\,n=2p+1}_{\textsc{fls}}$ does not contain non-vanishing finite terms.
Therefore the non-vanishing finite terms $G^{\hs(0)}_{\text{F, sing}}$ only need to be considered when $n$ is even.
\section{Singularity structure of $G^{}_{\mathrm{H}}$\label{ssgh}}
As will be seen in Sec.~\ref{hr}, expectation values obtained using HR are constructed from the regular, state-dependent object $W$.
Unlike the purely-geometric terms $U$ and $V$ in the Hadamard form \eqr{Hadx}, the quantity $W$ cannot be uniquely determined from recurrence relations similar to (\ref{e:Hcoeffs}).
Instead we may find $W$ using the Hadamard form \eqr{Hadx} and the explicit form of the Feynman Green's function
$\Gf$, given in (\ref{e:Gf2se}, \ref{e:Gfn2p+2}, \ref{e:Gfose}) for $n=2$, $n>2$ with $n$ even, and $n>2$ with $n$ odd respectively.

In this section we regularize the Hadamard form \eqr{Hadx} by splitting it into two parts,
\be
\Gfh=\Gfhr+\Gfhs,
\ee
where
\be
\Gfhr\isdef i\kappa W,
\ee
and
\be
\Gfhs\isdef i\kappa\left(U\sigma_{}^{1-\frac{n}{2}}+V\ln\bar{\sigma}\right) .
\ee
Given that the Feynman Green's function $\Gf$ and the Hadamard form $\Gfh$ are alternative expressions for the same object (as long as the
two space-time points are sufficiently close together),
the state-dependent quantity $W$ can then be determined using
\be \lble{Wdef}
W=- i\kappa^{-1}_{}\left(\Gf-\Gfhs\right),
\ee
since $\Gfhs$ is the purely geometric part responsible for the short-distance divergences. The computation of expressions for $\Gfhs$ for specific $n$ is discussed below.
\subsubsection{$n=2$\label{ghs2}}
In two space-time dimensions, the Hadamard form \eqr{Hadx} does not contain a geometric term $U$, so the singular part is simply:
\be \lble{Gh2}
G_{\text{H, sing}}^{n=2}\isdef\frac{ i}{2\pi}V\left(\ln\bar{s}-\frac{1}{2}\ln2\right).
\ee
\subsubsection{$n>2$, $n$ even\label{ghse}}
When $n>2$ is even, the singular part of the Hadamard form is
\be
G_{\text{H, sing}}^{n=2p+2}\isdef i\kappa\left[2_{}^{p}Us_{}^{-2p}+2V\left(\ln\bar{s}-\frac{1}{2}\ln2\right)\right].
\ee
As the functions $U$ and $V$ multiply poles and logarithmically divergent terms respectively, it is convenient to split $G_{\text{H, sing}}^{n=2p+2}$ into parts exhibiting these different behaviours.  Accordingly, we define
\begin{subequations}
\begin{align}
\lble{GHU}
G_{\text{H}(U)}^{n=2p+2}\isdef\;& i\kappa2_{}^{p}Us_{}^{-2p},\\
\lble{GHV}
G_{\text{H}(V)}^{n=2p+2}\isdef\;&2 i\kappa V\left(\ln\bar{s}-\frac{1}{2}\ln2\right).
\end{align}
\end{subequations}
\subsubsection{$n>2$, $n$ odd\label{ghso}}
When $n>2$ is odd, the geometric term $V$ is absent from the singular part of the Hadamard form, so we have:
\be
\lble{GHUodd}
G_{\text{H, sing}}^{n=2p+1}=G_{\text{H}(U)}^{n=2p+1}\isdef i\kappa2_{}^{p-\frac{1}{2}}Us_{}^{1-2p}.
\ee

\subsection{Computation of $G_{\mathrm{H}(U)}^{n}$}
For each $n$, the geometric quantity $U$ in (\ref{e:GHU}, \ref{e:GHUodd}) is found from the expansions (\ref{e:Usig}).
The coefficients $U^{}_{l}$ (with $l>0$) are found by exploiting \eqr{Hcoeffsa}, which can be written in the following form
since all quantities depend only on $s\xx$: %
\be \lble{RReg4s}
(l+1)U^{}_{l+1}+U^{}_{l+1,\hs s}\hs s-U^{}_{l+1}\Delta_{}^{-\frac{1}{2}}\Delta_{}^{\frac{1}{2}}{}_{,\hs s}^{}\hs s=\hat{U}^{}_{l},
\ee
with $l=0,1,2,\ldots,\frac{n}{2}-3$, where `${,\hs s}$' denotes a derivative with respect to the distance $s$,  and where
\be
\hat{U}^{}_{l}\isdef\left(\frac{\Box_{x}^{}-m^{2}_{\xi}}{n-2l-4} \right) U^{}_{l}.
\ee
Multiplying through by $\Delta^{-\frac{1}{2}}_{}s_{}^{l}$, the left-hand side of \eqr{RReg4s} can be written as $(U_{l+1}^{}\Delta^{-\frac{1}{2}}_{}s_{}^{l+1}){\hspace{-1mm}}^{}_{\;,\hs s}$.
The determination of each $U_{l}$ reduces to the following integration
\be
U^{}_{l+1}=K\frac{\Delta^{\frac{1}{2}}_{}}{s_{}^{l+1}}+\frac{\Delta^{\frac{1}{2}}_{}}{s_{}^{l+1}}\int_{0}^{s}\hat{U}^{}_{l}\hs\Delta_{}^{-\frac{1}{2}}{\acute{s}}{\hs}_{}^{l}d\acute{s},
\ee
where the dummy integration variable is denoted as $\acute{s}$.
As $s\to0$, the geometric quantity $U^{}_{l}$ (and therefore $\hat{U}^{}_{l}$) is regular but this is not generally so for $\Delta_{}^{\frac{1}{2}}s_{}^{-(l+1)}$, and so the constant of integration $K$ vanishes. Therefore
\be \lble{Uinte}
U^{}_{l+1}=\frac{\Delta_{}^{\frac{1}{2}}}{s_{}^{l+1}}\int_{0}^{s}\hat{U}^{}_{l}\Delta_{}^{-\frac{1}{2}}{\acute{s}}{\hs}_{}^{l}d\acute{s}.
\ee

From the explicit form of the Van-Vleck determinant on $adS^{}_{n}$ \eqr{VVDAdSn}, it can be seen that $\Delta $ and hence $U_{0}$ have Taylor series expansions in even powers of $s$.
By induction, using \eqr{Uinte}, this is true for all $U_{l}$, which we therefore write as:
\be \lble{Ulexp}
U_{l}=\sum_{j=0}^{+\infty}\tilde{u}^{}_{jl}s^{2j},
\ee
with expansion coefficients $\tilde{u}^{}_{jl}$.
The truncation of the expansion of $G_{\text{H}(U)}^{n}$ to second order in $s$ is then given by
\be \lble{GhUeII}
G_{\text{H}(U)}^{\hs\textbf{II},\,n}\isdef  i\kappa2_{}^{\frac{n}{2}-1}\sum_{l=0}^{p-1}\sum_{j=0}^{j_{p}}2_{}^{-l}\tilde{u}^{}_{jl}s_{}^{2j+2l-n+2},
\ee
where $n=2p+2$ if $n$ is even and $n=2p+1$ if $n$ is odd, and
\be
j_{p}=
\begin{cases}
p-l+1, & \qquad  n=2p+2,\qquad \\
p-l, & \qquad n=2p+1.
\end{cases}
\ee
If $n$ is even, the second-order truncation \eqr{GhUeII} possesses non-vanishing finite terms, $G_{\text{H, sing}}^{\hs(0),\,n=2p+2}$ corresponding to the $j=p-l$ summands.
As discussed at the end of Sec.~\ref{ssgf}, these non-vanishing finite terms will need to be considered when we compute renormalized expectation values for $n$ even, particularly as, in general,
\be \lble{frtneq}
G^{\hs(0)}_{\text{H, sing}}\neq G^{\hs(0)}_{\text{F, sing}}.
\ee

\subsection{Computation of $G_{\mathrm{H}(V)}^{n}$\label{GHseHV}}
This part of the Hadamard form is non-zero only when $n$ is even.
The function $V$ in (\ref{e:Gh2}, \ref{e:GHV}) is a solution of \eqr{sfwe} \cite{D&F08}, and therefore it may be expressed as a linear combination of the same hypergeometric functions present in \eqr{CamGS},
\begin{align}
V=\;&A\hf{\tfrac{n-1}{2}+\eta}{\tfrac{n-1}{2}-\eta}{\tfrac{n}{2}}{z}\nno\\
&\qquad+B\hf{\tfrac{n-1}{2}+\eta}{\tfrac{n-1}{2}-\eta}{\tfrac{n}{2}}{1-z},
\end{align}
where $A$ and $B$ are constants.
Since $V$ must be finite as $s\to0$, this implies that $B=0$, yielding
\be \lble{Vhyp}
V=A\hf{\tfrac{n-1}{2}+\eta}{\tfrac{n-1}{2}-\eta}{\tfrac{n}{2}}{z}.
\ee
Considering the expansion of $V$ in powers of $\sigma $ \eqr{Vsigb},
\be \lble{Alim}
A=\lim_{s\to0}V^{}_{0}.
\ee
The coefficient $V^{}_{0}$ is obtained from \eqr{HVBCb},
\be
\left(m_{\xi}^{2} - \Box_{x}^{}\right)U^{}_{\frac{n-4}{2}} =
(n-2)V^{}_{0}+2V^{}_{0,\hs s}\hs
-2V^{}_{0}\hs\Delta_{}^{-\frac{1}{2}}\Delta_{}^{\frac{1}{2}}{}_{,\hs s}^{}\hs s,
\ee
recast here in terms of $s$.
Therefore, since $V_{0}$ has a Taylor series expansion in even powers of $s$ only,
\be \lble{V0Adef}
A=\lim_{s\to0}\left(\frac{m^2_{\xi}-\Box_{x}^{}}{n-2}U_{\frac{n-4}{2}}\right).
\ee
Now that $A$ is fixed, the exact expression \eqr{Vhyp} can be easily used to find the expansion of $G_{\text{H}(V)}^{n}$ to the required order in $s$.

\section{Hadamard renormalization\label{hr}}

\subsection{Renormalized expectation values \label{fW}}

Having regularized the Feynman Green's function $\Gf$ and considered the singularity structure of the Hadamard form $\Gfh$, we are now in a position to find renormalized expectation values, in particular the RQFF and RSET.
Both these are determined from the state-dependent biscalar $W\xx $ appearing in the Hadamard form \eqr{Hadx}, which is calculated using \eqr{Wdef}.
In this paper we are interested in the global $adS^{}_{n}$ vacuum state $\vak$ which possesses all the symmetries of the underlying $adS^{}_{n}$ space-time.
Therefore the geometric quantities $U$ and $V$, and the Feynman Green's function \eqr{CamGS} depend only on the proper distance $s\xx$ between the two
space-time points $x$ and $x'$.
From \eqr{Wdef}, we deduce that, for the $adS^{}_{n}$ vacuum state the quantity $W$ also depends only on $s\xx $.
This simplifies our calculations greatly.

The simplest non-trivial expectation value is the RQFF, which is given by \cite{D&F08}:
\be \lble{psrdef}
\psr\isdef\kappa w,
\ee
where
\be \lble{w}
w\isdef \lim_{s\to 0}W.
\ee
The RSET is given by the action of a second-order linear differential operator $\boldsymbol{T}_{\!\mu\nu}^{}$ on $W$ \cite{D&F08}:
\be \lble{rsetdef}
\Tmnr\isdef\kappa\lim_{s\to0}\boldsymbol{T}_{\!\mu\nu}^{}W+\Theta^{}_{\mu\nu},
\ee
where $\boldsymbol{T}_{\!\mu\nu}^{}$ is a point-separated second-order linear differential operator version of the classical stress-energy tensor \cite{SMC}:
\begin{align} \lble{rsetop}
\boldsymbol{T}^{}_{\!\mu\nu}=\;&(1-2\xi){g^{}_{\nu}}_{}^{\nu'}\nabla^{}_{\mu}\nabla^{}_{\nu'}+\left(2\xi-\frac{1}{2}\right)g_{\mu\nu}^{} g_{}^{\rho\sigma'}\nabla^{}_{\rho}\nabla^{}_{\sigma'}\nno\\
&\qquad-2\xi {g^{}_{\mu}}_{}^{\mu'}{g^{}_{\nu}}_{}^{\nu'}\nabla^{}_{\mu'}\nabla^{}_{\nu'}+2\xi g_{\mu\nu}^{}\nabla^{}_{\rho}\nabla^{\rho}_{}\nno\\
&\qquad+\xi\left(\Rmn-\frac{1}{2}g_{\mu\nu}^{}R\right)-\frac{1}{2}g_{\mu\nu}^{} m_{}^{2},
\end{align}
where $\Rmn$ is the Ricci tensor.

The RSET satisfies Wald's axioms \cite{Wald2}, and hence is determined only up to the addition of a local, conserved, purely geometric, tensor $\Theta^{}_{\mu\nu}$ \cite{Wald2}.
As discussed in \cite{D&F08}, this ambiguity in the definition of $\Tmnr$ in curved space-times is related to the freedom of choice of renormalization mass scale $M$ present in HR, which is manifest in \eqr{bar}.
In HR, the local tensor $\Theta^{}_{\mu\nu}$ can be found explicitly in terms of the renormalization mass scale $M$ as follows \cite{D&F08}.

Firstly we note that, due to the maximal symmetry of the $adS^{}_{n}$ vacuum, in the limit $s\to 0$, the quantity $W$ will be a constant throughout space-time, and therefore the RQFF is also a constant.
With this in mind, the definition \eqr{rsetdef} can be reduced to \cite{D&F08}
\be \lble{rsetads}
\Tmnr=\kappa\left(-w^{}_{\mu\nu}+\xi\Rmn w-g^{}_{\mu\nu} v^{}_{1}\right)+\Theta^{}_{\mu\nu},
\ee
where
\be\lble{vbits}
w^{}_{\mu\nu}\isdef\lim_{s\to0}W^{}_{;\hs\mu\nu},
\qquad
v^{}_{l}\isdef\lim_{s\to0}V^{}_{l},
\quad
l=0,1,\ldots,
\ee
recalling from \eqr{Vsiga} that for $n$ odd, the term $v_{1}^{}$ will disappear.

For $n$ even, given \eqr{bar}, the Hadamard form \eqr{Hadx} can be written
\be\lble{Hadxamb}
\Gfh= i\kappa\left[U\sigma_{}^{1-\frac{n}{2}}+V\left(\ln M^{2}_{}+\ln\sigma\right)+W\right].
\ee
This implies that changing the renormalization mass scale from unity (corresponding to the Planck mass) to $M$ corresponds to the transformation
\be \lble{ambind}
W\to W+V\ln M^{2}_{}.
\ee
It follows therefore from \eqr{rsetdef} that the ambiguity in the definition of $\Tmnr$ is given by
\be
\Theta^{}_{\mu\nu}=-\kappa\lim_{s\to0}\boldsymbol{T}^{}_{\!\mu\nu}V\ln M^{2}_{},
\ee
which is zero when $n$ is odd, but non-vanishing when $n$ is even.
Since $V$ depends only on $s\xx $ \eqr{Vhyp}, the computation of $\Theta^{}_{\mu\nu}$ simplifies considerably \cite{D&F08}:
\be \lble{theads}
\Theta^{}_{\mu\nu}=\kappa\left[\left(v^{}_{0\hs\mu\nu}+g^{}_{\mu\nu} v^{}_{1}\right)-\xi\Rmn v^{}_{0}\right]\ln M^{2}_{}.
\ee
where
\be
\lble{v0mnhad}v^{}_{0\hs\mu\nu}\isdef\lim_{s\to0}V^{}_{0}{}^{}_{;\hs\mu\nu},
\ee
and $v^{}_{0}$ and $v^{}_{1}$ are given by \eqr{vbits}.

To find the RQFF \eqr{psrdef} and RSET \eqr{rsetads} we therefore need to compute $w$, $w^{}_{\mu\nu}$, $v^{}_{0}$, $v^{}_{0\hs \mu\nu }$ and $v_{1}$.
Since we have to take two derivatives of $w$ and $v^{}_{0}$, we require $W^{\textbf{II}}$ and $V^{\textbf{II}}$, respectively the expansions of
$W$ and $V$ to second order in the (small) separation $s$.
For the remainder of this section, we give the results for $W^{\textbf{II}}$ and $V^{\textbf{II}}$ for $n=2$ to $n=11$ inclusive, before calculating the
RQFF and RSET in the next section.

Firstly, $V^{\textbf{II}}$, the expansion of $V$, is found from the exact expression \eqr{Vhyp} and given in Sec.~\ref{VII}.
Finding $W^{\textbf{II}}$, the expansion of $W$, is more complicated.
The first step is, for each $n$, to find explicitly the expansions $G_{\text{F}}^{\hs\textbf{II},\,n}$ of the Feynman Green's function, using the general
expressions from Sec.~\ref{ssgf}.
In Sec.~\ref{frts} we present, for $n=2$ to $n=11$ inclusive, the zeroth order expansions $G_{\text{F, sing}}^{\hs(0)}$ to illustrate our results.  The expressions for the full second-order expansions $G_{\text{F}}^{\hs\textbf{II},\,n}$ are extremely lengthy so we do not present them here.
Next we find the second order expansions of the singular part of the Hadamard form, $G_{\text{H, sing}}^{\hs\textbf{II}}$ and then use them to find $W^{\textbf{II}}$
from \eqr{Wdef}, presenting our results in Sec.~\ref{WII}.
\vspace{-1mm}
\begin{widetext}
\subsection{Expressions for $V^{\textbf{II}}_{}$\label{VII}}
The second-order expansions of the geometric function $V$ appearing in the Hadamard form \eqr{Hadx} are required only when $n$ is even.
Using the results of Sec.~\ref{GHseHV}, we find the following:
\begin{subequations}
\label{e:VII}
\begin{align}
\lble{VII2}
V^{\textbf{II},\hs n=2}_{}=&-1-\left(\frac{1}{4}\eta_{}^{2}-\frac{1}{16}\right)\frac{s^{2}_{}}{{a}^{2}_{}},
\\
\lble{VII4}
V^{\textbf{II},\hs n=4}_{}=&
\frac{1}{a^{2}_{}}\left[\frac{1}{2}\eta_{}^{2}-\frac{1}{8}+\left(\frac{1}{16}\eta_{}^{4}-\frac{5}{32}\eta_{}^{2}+\frac{9}{256}\right)\frac{s^{2}_{}}{{a}^{2}_{}}\right],
\\
\lble{VII6}
V^{\textbf{II},\hs n=6}_{}=& \frac{1}{a^{4}_{}}\left[-\frac{1}{8}\eta_{}^{4}+\frac{5}{16}\eta_{}^{2}-\frac{9}{128}-\left(\frac{1}{96}\eta_{}^{6}-\frac{35}{384}\eta_{}^{4}+\frac{259}{1536}\eta_{}^{2}-\frac{75}{2048}\right)\frac{s^{2}_{}}{{a}^{2}_{}}\right],
\\
\lble{VII8}
V^{\textbf{II},\hs n=8}_{}=& \frac{1}{a^{6}_{}}\left[\frac{1}{96}\eta_{}^{6}-\frac{35}{384}\eta_{}^{4}+\frac{259}{1536}\eta_{}^{2}-\frac{75}{2048}+\left(\frac{1}{1536}\eta_{}^{8}-\frac{7}{512}\eta_{}^{6}+\frac{329}{4096}\eta_{}^{4}-\frac{3229}{24576}\eta_{}^{2}+\frac{3675}{131072}\right)\frac{s^{2}_{}}{{a}^{2}_{}}\right],
\\
V^{\textbf{II},\hs n=10}_{}=&
\frac{1}{a^{8}_{}}\left[-\frac{1}{2304}\eta_{}^{8}+\frac{7}{768}\eta_{}^{6}-\frac{329}{6144}\eta_{}^{4}+\frac{3229}{36864}\eta_{}^{2}-\frac{1225}{65536}\right.\nno \\
& \qquad -\left.\left(\frac{1}{46080}\eta_{}^{10}-\frac{11}{12288}\eta_{}^{8}+\frac{1463}{122880}\eta_{}^{6}-\frac{17281}{294912}\eta_{}^{4}+\frac{117469}{1310720}\eta_{}^{2}-\frac{19845}{1048576}\right)\frac{s^{2}_{}}{{a}^{2}_{}}\right].
\lble{VII10}
\end{align}
\end{subequations}

\subsection{Expansions of the Feynman Green's function\label{frts}}
The singularity structure of the Feynman Green's function has been discussed in detail in Sec.~\ref{ssgf}.
Here we present our results for the zeroth order expansions $G_{\text{F, sing}}^{\hs(0)}$, which are sufficient for the computation of the RQFF.
The second order expansions $G_{\text{F}}^{\hs\textbf{II},\,n}$, required for the computation of the RSET, are computed using the method of Sec.~\ref{ssgf}, but the resulting expressions are too lengthy to reproduce here.
The explicit forms for the zeroth order expansions are:
\begin{subequations}
\label{e:Gfs}
\begin{align}
\lble{Gfs2}
G_{\text{F, sing}}^{\hs\boldsymbol{0},\,n=2}=\;&-\frac{ i}{2\pi}\ln\bar{s},\\
\lble{Gfs3}
G_{\text{F, sing}}^{\hs\boldsymbol{0},\,n=3}=\;&\frac{ i}{4\pi s},\\
\lble{Gfs4}
G_{\text{F, sing}}^{\hs\boldsymbol{0},\,n=4}=\;&\frac{ i}{4\pi^{2}}\left\{\frac{1}{s^{2}}+\left[\frac{m^{2}}{2}+\left(1-6\xi\right)\frac{1}{a^{2}}\right]\ln\bar{s}-\frac{1}{12a^{2}}\right\},\\
\lble{Gfs5}
G_{\text{F, sing}}^{\hs\boldsymbol{0},\,n=5}=\;&\frac{ i}{8\pi^{2}}\left\{\frac{1}{s^{3}}-\left[\frac{m^{2}_{}}{2}+\left(2-10\xi\right)\frac{1}{a^{2}}\right]\frac{1}{s}\right\},
\\
\lble{Gfs6}
G_{\text{F, sing}}^{\hs\boldsymbol{0},\,n=6}=\;&\frac{ i}{4\pi^{3}}\left\{\frac{1}{s^{4}}-\left[\frac{m^{2}}{4}+\left(\frac{5}{3}-\frac{15}{2}\xi\right)\frac{1}{a^{2}_{}}\right]\frac{1}{s^{2}}\right.\nno\\
&\qquad\left.-\left[\frac{m^{4}}{16}+\left(\frac{5}{8}-\frac{15}{4}\xi\right)\frac{m^{2}}{a^{2}}+\left(\frac{3}{2}-\frac{75}{4}\xi+\frac{225}{4}\xi^{2}\right)\frac{1}{a^{4}}\right]\ln\bar{s}+\frac{m^{2}}{48a^{2}}+\left(\frac{101}{720}-\frac{5}{8}\xi\right)\frac{1}{a^{4}}\right\},\\
\lble{Gfs7}
G_{\text{F, sing}}^{\hs\boldsymbol{0},\,n=7}=\;&\frac{3 i}{16\pi^{3}}\left\{\frac{1}{s^{5}}\left[\frac{m^{2}}{6}+\left(\frac{5}{3}-{7}\xi\right)\frac{1}{a^{2}}\right]+\frac{1}{s^{3}}\left[\frac{m^{4}}{24}+\left(\frac{2}{3}-\frac{7}{2}\xi\right)\frac{m^{2}}{a^{2}}+\left(\frac{8}{3}-28\xi+\frac{147}{2}\xi^{2}\right)\frac{1}{a^{4}}\right]\frac{1}{s}\right\},\\
\lble{Gfs8}
G_{\text{F, sing}}^{\hs\boldsymbol{0},\,n=8}=\;&\frac{ i}{2\pi^{4}}\left\{\frac{1}{s^{6}}-\left[\frac{m^{2}}{8}+\left(\frac{7}{4}-7\xi\right)\frac{1}{a^{2}}\right]\frac{1}{s^{4}}+\left[\frac{m^{4}}{64}+\left(\frac{35}{96}-\frac{7}{4}\xi\right)\frac{m^{2}}{a^{2}}+\left(\frac{259}{120}-\frac{245}{12}\xi+49\xi^{2}\right)\frac{1}{a^{4}}\right]\frac{1}{s^{2}}\right.\nno\\
&\qquad +\left[\frac{m^{6}}{384}+\left(\frac{7}{96}-\frac{7}{16}\xi\right)\frac{m^{4}}{a^{2}}+\left(\frac{21}{32}-\frac{49}{6}\xi+\frac{49}{2}\xi^{2}\right)\frac{m^{2}}{a^{4}}+\left(\frac{15}{8}-\frac{147}{4}\xi+\frac{686}{3}\xi^{2}-\frac{1372}{3}\xi^{3}\right)\frac{1}{a^{6}}\right]\ln\bar{s}\nno\\
&\qquad \left.-\frac{m^{4}}{768a^{2}}-\left(\frac{11}{360}-\frac{7}{48}\xi\right)\frac{m^{2}}{a^{4}}-\left(\frac{11027}{60480}-\frac{77}{45}\xi+\frac{49}{12}\xi^{2}\right)\frac{1}{a^{6}}\right\},\\
\lble{Gfs9}
G_{\text{F, sing}}^{\hs\boldsymbol{0},\,n=9}=\;&\frac{15 i}{32\pi^{4}}\left\{\frac{1}{s^{7}}-\left[\frac{m^{2}}{10}
+\left(\frac{28}{15}-\frac{36}{5}\xi\right)\frac{1}{a^{2}}\right]\frac{1}{s^{5}}+\left[\frac{m^{4}}{120}+\left(\frac{4}{15}
-\frac{6}{5}\xi\right)\frac{m^{2}}{a^{2}}+\left(\frac{98}{45}-\frac{96}{5}\xi+\frac{216}{5}\xi^{2}\right)\frac{1}{a^{4}}\right]
\frac{1}{s^{3}}\right.\nno\\
&\qquad \left.-\left[\frac{m^{6}}{720}+\left(\frac{1}{18}-\frac{3}{10}\xi\right)\frac{m^{4}}{a^{2}}+\left(\frac{11}{15}-
8\xi+\frac{108}{5}\xi^{2}\right)\frac{m^{2}}{a^{4}}
+\left(\frac{16}{5}-\frac{264}{5}\xi+288\xi^{2}-\frac{2592}{5}\xi^{3}\right)\frac{1}{a^{6}}\right]\frac{1}{s}\right\},\\
\lble{Gfs10}
G_{\text{F, sing}}^{\hs\boldsymbol{0},\,n=10}=\;&\frac{3 i}{2\pi^{5}}\left\{\frac{1}{s^{8}}-\left[\frac{m^{2}}{12}+\left(2-\frac{15}{2}\xi\right)\frac{1}{a^{2}}\right]\frac{1}{s^{6}}+\left[\frac{m^{4}}{192}+\left(\frac{7}{32}-\frac{15}{16}\xi\right)\frac{m^{2}}{a^{2}}+\left(\frac{47}{20}-\frac{315}{16}\xi+\frac{675}{16}\xi^{2}\right)\frac{1}{a^{4}}\right]\frac{1}{s^{4}}\right.\nno\\
&\qquad-\left[\frac{m^{6}}{2304}+\left(\frac{3}{128}-\frac{15}{128}\xi\right)\frac{m^{4}}{a^{2}}+\left(\frac{203}{480}-\frac{135}{32}\xi+\frac{675}{64}\xi^{2}\right)\frac{m^{2}}{a^{4}}
+\left(\frac{3229}{1260}-\frac{609}{16}\xi+\frac{6075}{32}\xi^{2}
\right. \right.\nno \\
& \qquad \left. \left.-\frac{10125}{32}\xi^{3}\right)\frac{1}{a^{6}}\right]\frac{1}{s^{2}}
-\left[\frac{1}{18432}m^{8}+\left(\frac{5}{1536}-\frac{5}{256}\xi\right)\frac{m^{6}}{a^{2}}+\left(\frac{109}{1536}-\frac{225}{256}\xi+\frac{675}{256}\xi^{2}\right)\frac{m^{4}}{a^{4}}\right.\nno\\
&\qquad\left.+\left(\frac{761}{1152}-\frac{1635}{128}\xi+\frac{10125}{128}\xi^{2}-\frac{10125}{64}\xi^{3}\right)\frac{m^{2}}{a^{6}}
\left.+\left(\frac{35}{16}-\frac{3805}{64}\xi+\frac{73575}{128}\xi^{2}-\frac{151875}{64}\xi^{3}\right. \right. \right.
\nno \\ & \qquad \left. \left. \left.
+\frac{455625}{128}\xi^{4}\right)\frac{1}{a^{8}}\right]\ln\bar{s}
+\frac{m^{6}}{27648a^{2}}+\left(\frac{271}{138240}-\frac{5}{512}\xi\right)\frac{m^{4}}{a^{4}}+\left(\frac{51601}{1451520}-\frac{271}{768}\xi+\frac{225}{256}\xi^{2}\right)\frac{m^{2}}{a^{6}}\right.\nno\\
&\qquad\left.+\left(\frac{262349}{1209600}-\frac{51601}{16128}\xi+\frac{4065}{256}\xi^{2}-\frac{3375}{128}\xi^{3}\right)\frac{1}{a^{8}}\right\},
\end{align}
\newpage
\begin{align}
\lble{Gfs11}
G_{\text{F, sing}}^{\hs\boldsymbol{0},\,n=11}=\;&\frac{105 i}{64\pi^{5}}\left\{\frac{1}{s^{9}}-\left[\frac{m^{2}}{14}+\left(\frac{15}{7}-\frac{55}{7}\xi\right)\frac{1}{a^{2}}\right]\frac{1}{s^{7}}+\left[\frac{m^{4}}{280}+\left(\frac{4}{21}-\frac{11}{14}\xi\right)\frac{m^{2}}{a^{2}}+\left(\frac{13}{5}-\frac{440}{21}\xi+\frac{605}{14}\xi^{2}\right)\frac{1}{a^{4}}\right]\frac{1}{s^{5}}\right.\nno\\
&\qquad-\left[\frac{m^{6}}{5040}+\left(\frac{1}{72}-\frac{11}{168}\xi\right)\frac{m^{4}}{a^{2}}+\left(\frac{103}{315}-\frac{55}{18}\xi+\frac{605}{84}\xi^{2}\right)\frac{m^{2}}{a^{4}}\right. \nno \\
& \qquad \left.\left.+\left(\frac{164}{63}-\frac{2266}{63}\xi+\frac{3025}{18}\xi^{2}-\frac{33275}{126}\xi^{3}\right)\frac{1}{a^{6}}\right]\frac{1}{s^{3}}\right.
+\left[\frac{m^{8}}{40320}+\left(\frac{1}{504}-\frac{11}{1008}\xi\right)\frac{m^{6}}{a^{2}} \right. \nno \\
& \qquad \left.
+\left(\frac{37}{630}-\frac{55}{84}\xi+\frac{605}{336}\xi^{2}\right)\frac{m^{4}}{a^{4}}+\left(\frac{16}{21}-\frac{814}{63}\xi+\frac{3025}{42}\xi^{2}-\frac{33275}{252}\xi^{3}\right)\frac{m^{2}}{a^{6}}\right.\nno\\
&\qquad\left.\left.+\left(\frac{128}{35}-\frac{1760}{21}\xi+\frac{44770}{63}\xi^{2}-\frac{166375}{63}\xi^{3}+\frac{1830125}{504}\xi^{4}\right)\frac{1}{a^{8}}\right]\frac{1}{s}\right\}.
\end{align}
\end{subequations}
\end{widetext}

The corresponding expansions of the singular part of the Hadamard form $\Gfhs$ can be found using the method of Sec.~\ref{ssgh}.  We do not give the resulting expressions explicitly here.
The form of $\Gfhs$ for a general space-time with $n=2$ to $n=6$ dimensions can be found in \cite{D&F08}.
Typically, such expressions involve biscalar functions and are ridden with Riemann tensor polynomials.  However, the maximal symmetry of $adS^{}_{n}$ simplifies these expressions by involving scalar functions only and reducing the possible types of Riemann tensor polynomials significantly. The singular terms of the results \eqr{Gfs} agree with their counterparts in \cite{D&F08} for the case of $adS^{}_{n}$.

For $n$ odd, we find that
\be
G_{\text{F, sing}}^{\hs(0)}-G_{\text{H, sing}}^{\hs(0)}=0,
\ee
so that the zeroth order expansion of the singular part of the Hadamard form equals that of the Feynman Green's function given in (\ref{e:Gfs}).
For $n$ even, the singular terms in $G_{\text{F, sing}}^{\hs(0)}$ and $G_{\text{H, sing}}^{\hs(0)}$ match, but there are non-vanishing finite renormalization terms (FRTs), defined by
\be
\lble{frt}
G_{\text{F, \textsc{frt}}}^{}\isdef\;\lim_{s\to0}\left(\Gfs-\Gfhs\right)
=\;G_{\text{F, sing}}^{\hs(0)}-G_{\text{H, sing}}^{\hs(0)}.
\ee
\begin{widetext}
Explicit expressions for the FRTs for $n=2,4,6,8$ and $10$ are:
\begin{subequations}
\begin{align}
\lble{fr2}
G_{\text{F, \textsc{frt}}}^{\hs n=2}=\;&\frac{ i}{4\pi}\ln2, \\
\lble{fr4}
G_{\text{F, \textsc{frt}}}^{\hs n=4}=\;&\frac{ i}{4\pi^{2}}\left\{\frac{1}{6a^{2}}+2\left[\frac{m^{2}}{2}+\left(1-6\xi\right)\frac{1}{a^{2}}\right]\ln2\right\}, \\
\lble{fr6}
G_{\text{F, \textsc{frt}}}^{\hs n=6}=\;&\frac{ i}{4\pi^{3}}\left\{-\frac {m^{2}}{{12a}^{2}}-\left(\frac{221}{480}-\frac{5}{2}\xi\right)\frac{1}{a^{4}}-4\left[\frac{m^{4}}{16}+\left(\frac{5}{8}-\frac{15}{4}\xi\right)\frac{m^{2}}{a^{2}}+\left(\frac{3}{2}-\frac{75}{4}\xi+\frac{225}{4}\xi^{2}\right)\frac{1}{a^{4}}\right]\ln2\right\},
\\
\lble{fr8}
G_{\text{F, \textsc{frt}}}^{\hs n=8}=\;&\frac{ i}{2\pi^{4}}\left\{{\frac{m^{4}}{128a^{2}}}+\left(\frac{1811}{11520}-\frac{7}{8}\xi\right)\frac{m^{2}}{a^{4}}+\left(\frac{46639}{60480}-\frac{12677}{1440}\xi+\frac {49}{2}\xi^{2}\right)\frac{1}{a^{6}}\right.\nno\\
&\qquad\left.+8\left[\frac{m^{6}}{384}+\left(\frac{7}{96}-\frac{7}{16}\xi\right)\frac{m^{4}}{a^{2}}+\left(\frac{21}{32}-\frac{49}{6}\xi+\frac{49}{2}\xi^{2}\right)\frac{m^{2}}{a^{4}}+\left(\frac{15}{8}-\frac{147}{4}\xi+\frac{686}{3}\xi^{2}-\frac{1372}{3}\xi^{3}\right)\frac{1}{a^{6}}\right]\ln2\right\},
\\
\lble{Gfrt10}
G_{\text{F, \textsc{frt}}}^{\hs n=10}=\;&\frac{3 i}{2\pi^{5}}\left\{-\frac{m^{6}}{3456a^{2}}-\left(\frac{3823}{276480}-\frac{5}{64}\xi\right)\frac{m^{4}}{a^{4}}-\left(\frac{624013}{2903040}-\frac{3823}{1536}\xi+\frac{225}{32}\xi^{2}\right)\frac{m^{2}}{a^{6}}\right.\nno\\
&\qquad -\left(\frac{2091139}{1935360}-\frac{624013}{32256}\xi+\frac{57345}{512}\xi^{2} -\frac{3375}{16}\xi^{3}\right)\frac{1}{a^{8}}-16\left[\frac{1}{18432}m^{8}+\left(\frac{5}{1536}-\frac{5}{256}\xi\right)\frac{m^{6}}{a^{2}}\right.\nno\\
&\qquad+\left(\frac{109}{1536}-\frac{225}{256}\xi+\frac{675}{256}\xi^{2}\right)\frac{m^{4}}{a^{4}}+\left(\frac{761}{1152}-\frac{1635}{128}\xi+\frac{10125}{128}\xi^{2}-\frac{10125}{64}\xi^{3}\right)\frac{m^{2}}{a^{6}}\nno \\
& \qquad \left. \left. +\left(\frac{35}{16}-\frac{3805}{64}\xi+\frac{73575}{128}\xi^{2}-\frac{151875}{64}\xi^{3}+\frac{455625}{128}\xi^{4}\right)\frac{1}{a^{8}}\right]\ln2\right\}.
\end{align}
\end{subequations}
\newpage
\subsection{Expressions for $W^{\textbf{II}}_{}$\label{WII}}
Expressions for $W^{\textbf{II}}_{}$ are computed using results for $G^{\textbf{II}}_{\text{F}}$ and $G^{\textbf{II}}_{\text{H, sing}}$ obtained previously. and are given below for $n=2$ to $n=11$ inclusive as functions of $\eta$:
\begin{subequations}
\label{e:WII}
\begin{align}
\lble{WII2}
W^{\textbf{II},\hs n=2}_{}=\;&\frac{1}{a^{2}}\left\{  \left[ -2a_{}^{2}-\left(\frac{1}{2}\eta_{}^{2}-\frac{1}{8}\right) {s}^{2} \right]\!\mathit{\widetilde{\Upsilon\hskip1.2pt}}+ \left( \frac{1}{2}\eta_{}^{2}+\frac{1}{24}\right) {s}^{2} \right\},
\\
\lble{WII3}
W^{\textbf{II},\hs n=3}_{}=\;&\frac{\sqrt{2}}{a^{3}_{}}\left[ -\eta a_{}^{2}-\left(\frac{1}{6}\eta_{}^{3}-\frac{1}{6}\eta \right) {s}^{2} \right],
\\
\lble{WII4}
W^{\textbf{II},\hs n=4}_{}=\;&\frac{1}{a^{4}_{}}\left\{\left[\left(\eta_{}^{2}-\frac{1}{4}\right)a_{}^{2}+\left( \frac{1}{8}\eta_{}^{4}-\frac{5}{16}\eta_{}^{2}+\frac{9}{128}\right)s_{}^{2}\right]\!\mathit{\widetilde{\Upsilon\hskip1.2pt}}
-\left(\frac{1}{2}\eta_{}^{2}+\frac{1}{24}\right)a_{}^{2}-\left(\frac{5}{32}\eta_{}^{4}-\frac{35}{192}\eta_{}^{2}-\frac{47}{2560}\right)s_{}^{2}\right\},
\\
\lble{WII5}
W^{\textbf{II},\hs n=5}_{}=\;&\frac{\sqrt{2}}{a^{5}_{}}\left[ \left( \frac{2}{3}{\eta
}^{3}-\frac{2}{3}\eta \right) a_{}^{2}+ \left( \frac{1}{15}\eta_{}^{5}-\frac{1}{3}\eta_{}^{3}+
{\frac{4}{15}}\eta \right) {s}^{2} \right],
\\
\lble{WII6}
W^{\textbf{II},\hs n=6}_{}=\;&\frac{1}{a^{6}_{}}\left\{\left[\left(-\frac{1}{4}\eta_{}^{4}+\frac{5}{8}\eta_{}^{2}-{\frac{9}{64}}\right)a_{}^{2}-\left(\frac{1}{48}\eta_{}^{6}-\frac {35}{192}\eta_{}^{4}+\frac {259}{768}\eta_{}^{2}-\frac{75}{1024}\right)s^{2}_{}\right]\!\mathit{\widetilde{\Upsilon\hskip1.2pt}}\right.\nno\\
&\qquad+\left.\left(\frac{3}{16}\eta_{}^{4}-\frac{29}{96}\eta_{}^{2}-\frac{107}{3840}\right)a_{}^{2}+\left(\frac{17}{576}\eta_{}^{6}-\frac{403}{2304}\eta_{}^{4}+\frac{2741}{15360}\eta_{}^{2}+\frac{14171}{774144}\right)s^{2}_{}\right\},
\\
\lble{WII7}
W^{\textbf{II},\hs n=7}_{}=\;&\frac{\sqrt{2}}{a^{7}_{}}\left[\left(-\frac{4}{45}\eta_{}^{5}+\frac{4}{9}\eta_{}^{3}-\frac{16}{45}\eta\right)a_{}^{2}-\left(\frac{2}{315}\eta_{}^{7}-\frac{4}{45}\eta_{}^{5}+\frac{14}{45}\eta_{}^{3}-\frac{8}{35}\eta\right)s^{2}_{}\right],
\\
\lble{WII8}
W^{\textbf{II},\hs n=8}_{}=\;&\frac{1}{a^{8}_{}}\left\{ \left[ \left( \frac{1}{48}\eta_{}^{6}-{
\frac {35}{192}}\eta_{}^{4}+{\frac {259}{768}}\eta_{}^{2}-{\frac {75}{
1024}} \right) a_{}^{2}
+\left( {\frac {1}{768}}\eta_{}^{8}-{\frac {7}{
256}}\eta_{}^{6}+{\frac {329}{2048}}\eta_{}^{4}-{\frac {3229}{12288}}
\eta_{}^{2}+{\frac {3675}{65536}} \right) {s}^{2} \right]\!\mathit{\widetilde{\Upsilon\hskip1.2pt}}\right. \nno\\
&\qquad-\left(\frac{11}{576}\eta_{}^{6}-\frac{313}{2304}\eta_{}^{4}+\frac{2471}{15360}\eta_{}^{2}+\frac {11969}{774144}\right) a_{}^{2}\nno\\
&\qquad-\left.\left(\frac{37}{18432}\eta_{}^{8}-\frac{601}{18432}\eta_{}^{6}+\frac{103523}{737280}\eta_{}^{4}-\frac{4138441}{30965760}\eta^{2}_{}-\frac{195425}{14155776}\right)s^{2}_{}\right\},
\\
\lble{WII9}
W^{\textbf{II},\hs n=9}_{}=\;&\frac{\sqrt{2}}{a^{9}_{}}\left[ \left( {\frac {8
}{1575}}\eta_{}^{7}-{\frac {16}{225}}\eta_{}^{5}+{\frac {56}{225}}{
\eta}^{3}-{\frac {32}{175}}\eta \right) a_{}^{2}+ \left( {\frac {4}{
14175}}\eta_{}^{9}-{\frac {8}{945}}\eta_{}^{7}+{\frac {52}{675}}{\eta
}^{5}-{\frac {656}{2835}}\eta_{}^{3}+{\frac {256}{1575}}\eta \right)
{s}^{2} \right],
\\
\lble{WII10}
W^{\textbf{II},\hs n=10}_{}=\;&\frac{1}{a^{10}_{}}\left\{  \left[  \left( -{\frac {1}{1152}}
\eta_{}^{8}+{\frac {7}{384}}\eta_{}^{6}-{\frac {329}{3072}}\eta_{}^{4}+{
\frac {3229}{18432}}\eta_{}^{2}-{\frac {1225}{32768}} \right) a_{}^{2}\right.\right.\nno\\
&\qquad-\left.\left( {\frac {1}{23040}}\eta_{}^{10}-{\frac {11}{6144}}\eta_{}^{8}+
{\frac {1463}{61440}}\eta_{}^{6}-{\frac {17281}{147456}}\eta_{}^{4}+{
\frac {117469}{655360}}\eta_{}^{2}-{\frac {19845}{524288}} \right) {s}
^{2} \right]\!\mathit{\widetilde{\Upsilon\hskip1.2pt}}\nno
\\
&\qquad+ \left( {\frac {25}{27648}}\eta_{}^{8}-{\frac {461}{
27648}}\eta_{}^{6}+{\frac {87983}{1105920}}\eta_{}^{4}-{\frac {3854941
}{46448640}}\eta_{}^{2}-{\frac {288563}{35389440}} \right) a_{}^{2}\nno \\
&\qquad+\left.\left( {\frac {197}{2764800}}\eta_{}^{10}-{\frac {5381}{2211840}}{
\eta}^{8}+{\frac {579833}{22118400}}\eta_{}^{6}-{\frac {36356041}{
371589120}}\eta_{}^{4}+{\frac {442096687}{4954521600}}\eta_{}^{2}+{
\frac {19213927}{2076180480}} \right) {s}^{2} \right\},
\\
\lble{WII11}
W^{\textbf{II},\hs n=11}_{}=\;&\frac{\sqrt{2}}{a^{11}_{}}\left[  \left( -{\frac
{16}{99225}}\eta_{}^{9}+{\frac {32}{6615}}\eta_{}^{7}-{\frac {208}{
4725}}\eta_{}^{5}+{\frac {2624}{19845}}\eta_{}^{3}-{\frac {1024}{11025
}}\eta \right) a_{}^{2}\right.\nno\\
&\qquad-\left.\left( {\frac {8}{1091475}}\eta_{}^{11}-{
\frac {8}{19845}}\eta_{}^{9}+{\frac {248}{33075}}\eta_{}^{7}-{\frac {
1112}{19845}}\eta_{}^{5}+{\frac {15328}{99225}}\eta_{}^{3}-{\frac {512
}{4851}}\eta \right) {s}^{2} \right],
\end{align}
\end{subequations}
where (recalling \eqr{ups}), for even $n$,
\be
\lble{upstilde}
\mathit{\widetilde{\Upsilon\hskip1.2pt}}\isdef\psf{\tfrac{1}{2}+\eta}+\emc-\frac{1}{2}\ln2-\ln \bar{a}=\mathit{\Upsilon}+\frac{1}{2}\ln2.
\ee
\end{widetext}
\newpage

\section{Results\label{res}}
In this section we present our results for the RQFF in Sec.~\ref{respsr} and the RSET in Sec.~\ref{resrset}, calculated using the methodology developed in previous sections.  For each object, we begin by listing the algebraic expressions that have been obtained for $n=2$ to $n=11$ inclusive.  For both the RQFF and the RSET, these expressions are polynomials in the quantity $\eta$ (given by \eqr{eta}).  Since a significant part of the subsequent discussion of these expressions considers the behaviour of these objects as functions of $\eta$, we begin by repeating the definition \eqr{eta} here for convenience:
\be\lble{eta2}
\eta=\sqrt{m^{2}_{}a^{2}_{}-\xi n(n-1)+\frac{(n-1)^{2}_{}}{4}},
\ee
where we have made use of the fact that the Ricci scalar on $adS^{}_{n}$ is given by
\be\lble{RicADS}
R=-a^{-2}_{}n(n-1).
\ee

On the fixed background $adS^{}_{n}$ geometry, $\eta $ depends on both $m^{2}_{}$, the mass-squared of the scalar field \emph{and} $\xi$, its coupling to the background curvature. In our analysis we have assumed that $\eta $ is real, which implies an upper bound on $\xi $:
\be\lble{ximax}
\xi\leq\xi^{}_{\text{max}}\isdef\frac{m^{2}_{}a^{2}_{}}{n(n-1)}+\frac{n-1}{4n}.
\ee
It follows that an increase in $\eta$ is, in general, the result of a decrease in $\xi$ from $\xi^{}_{\text{max}}$ and/or an increase in $m^{2}_{}\ge 0$.
In the case of the RSET, the algebraic expressions for the object carry an explicit linear dependence on $\xi$, as well as their dependence on $\xi$ (and $m^{2}_{}$) through powers of $\eta $.

Our discussions of the results for the RQFF and the RSET are structured as follows.  We begin by commenting on the general form of the expressions for each object.  We then fix the renormalization mass scale $M$ and compare the resulting profiles for each number of space-time dimensions $n$.  Next, we plot this behaviour for even $n$ separately for a discrete selection of values of $M$. In all our analysis below, we fix the $adS^{}_{n}$ radius of curvature $a=1$ for convenience.

We emphasize that the algebraic expressions presented in Sec.~\ref{respsr} and Sec.~\ref{resrset} are valid for \emph{all} $\eta\geq0$ (and therefore for all values of $\xi$ in the range \eqr{ximax}).
In view of the additional explicit dependence of the RSET on the curvature-coupling $\xi$, plots such as those just described (separately highlighting the dependence on $n$ and $M$), require the value of $\xi$ to also be fixed.  For this purpose, we highlight the minimally-coupled case $\xi=0$ and the conformally-coupled case,
\be\lble{mc}
\xi=\frac{n-2}{4(n-1)},
\ee
when
\be
\lble{etac}
\eta = {\sqrt {m^{2}_{}a^{2}_{} + \frac {1}{4}}}.
\ee
In this article, these values are chosen for their simplifying effect on the expressions rather than to illustrate any special physical effects.  We have therefore included surface plots for each $n$.  In these plots, the dependence of the RSET on $\xi$ can be observed separately from the dependence on $\eta$.

\begin{widetext}
\subsection{Results for $\psr$\label{respsr}}
Using (\ref{e:psrdef}, \ref{e:WII}), expressions for $\psr$ are given below for $n=2$ to $n=11$ respectively (with ${\mathit{\widetilde{\Upsilon\hskip1.2pt}}}$ given by \eqr{ups}):
\begin{subequations}
\label{e:psr}
\begin{align}
\lble{psr2}
\psr^{n=2}=&\;-\frac{1}{2\pi}\mathit{\widetilde{\Upsilon\hskip1.2pt}},
\\
\lble{psr3}
\psr^{n=3}=&\;-\frac{1}{4\pi a}\,\eta,
\\
\lble{psr4}
\psr^{n=4}=&\;\frac{1}{8\pi^{2}a^{2}}\left[\left(\eta^{2}-\frac{1}{4}\right)\!\mathit{\widetilde{\Upsilon\hskip1.2pt}}-\frac{1}{2}\eta^{2}-\frac{1}{24}\right],
\\
\lble{psr5}
\psr^{n=5}=&\;\frac{1}{24\pi^{2}a^{3}}\left(\eta^{3}-\eta\right),
\\
\lble{psr6}
\psr^{n=6}=&\;-\frac{1}{64\pi^{3}a^{4}}\left[\left(\eta^{4}-\frac{5}{2}\eta^{2}+\frac{9}{16}\right)\!\mathit{\widetilde{\Upsilon\hskip1.2pt}}-\frac{3}{4}\eta^{4}+\frac{29}{24}\eta^{2}+\frac{107}{960}\right],
\\
\lble{psr7}
\psr^{n=7}=&\;-\frac{1}{240\pi^{3}a^{5}}\left(\eta^{5}-5\eta^{3}+4\eta\right),
\\
\lble{psr8}
\psr^{n=8}=&\;\frac{1}{768\pi^{4}a^{6}}\left[\left(\eta^{6}-\frac{35}{4}\eta^{4}+\frac{259}{16}\eta^{2}-\frac{225}{64}\right)\!\mathit{\widetilde{\Upsilon\hskip1.2pt}}-\frac{11}{12}\eta^{6}+\frac{313}{48}\eta^{4}-\frac{2471}{320}\eta^{2}-\frac{11969}{16128}\right],
\end{align}
\begin{align}
\lble{psr9}
\psr^{n=9}=&\;\frac{1}{3360\pi^{4}a^{7}}\left(\eta^{7}-14\eta^{5}+49\eta^{3}-36\eta\right),
\\
\lble{psr10}
\psr^{n=10}=&\;-\frac{1}{12288\pi^{5}a^{8}}\left[\left(\eta^{8}-21\eta^{6}+\frac{987}{8}\eta^{4}-\frac{3229}{16}\eta^{2}+\frac{11025}{256}\right)\!\mathit{\widetilde{\Upsilon\hskip1.2pt}}
\right.\nno\\&\qquad\left.
-\frac{25}{24}\eta^{8}+\frac{461}{24}\eta^{6}-\frac{87983}{960}\eta^{4}+\frac{3854941}{40320}\eta^{2}+\frac{288563}{30720}\right],
\\
\lble{psr11}
\psr^{n=11}=&\;-\frac{1}{60480\pi^{5}a^{9}}\left(\eta^{9}-30\eta^{7}+273\eta^{5}-820\eta^{3}+576\eta\right).
\end{align}
\end{subequations}
\end{widetext}
\subsubsection{General remarks}
\label{sec:phi2gen}
For each number of space-time dimensions $n$, the expressions \eqr{psr} depend on: the length scale $a$ (which determines the $adS^{}_{n}$ radius of curvature);  the mass of the field quanta $m$; and $\xi $, the coupling to the space-time curvature. The dependence on $m$ and $\xi $ is only via the constant $\eta$, given by \eqr{eta2}. For even $n$, there is an additional dependence on the renormalization mass scale $M$.  Once these parameters are fixed, the vacuum RQFF is a constant throughout $adS^{}_{n}$ for each $n$.

The results \eqr{psr} agree with expressions computed using zeta-function regularization \cite{Cal}, up to the addition of a constant. This difference may be attributed to the freedom in the definition of the renormalization mass scale, where Caldarelli's scale $M^{}_{\text{C}}$ is related to ours via
\be \lble{Mren}
M^{}_{\text{C}}=\sqrt{2}e_{}^{-\emc}M,
\ee
which agrees with that in \cite{Mor}.

The results \eqr{psr} are plotted in Fig.~\ref{f:p} as functions of $\eta$, in which we have set $a=1$ and $M=2^{-\frac{1}{2}}_{}e^{\emc}_{}$ (so that $M^{}_{\text{C}}=1$).
When $\eta=0$, the RQFF vanishes for $n$ odd, but for even $n$, the $\mathit{\widetilde{\Upsilon\hskip1.2pt}}$ factor present means that $\psr\neq0$.
The curves associated with consecutive pairs of consecutive values of $n$ alternate as to whether they are ultimately decreasing or increasing functions of $\eta$.
This behaviour arises from the mathematical structure of the Green's function, as exemplified by the factors of $(-1)^{p}$ in the formal Laurent series \eqr{Se} and \eqr{So} for space-times with $2p+2$ and $2p+1$ dimensions respectively. This overall factor of $(-1)^{p}$ multiplies the whole of the RQFF \eqr{psr}, and hence results in the large $\eta $ behaviour shown in Fig.~\ref{f:p}.

\begin{figure}[H]
\begin{center}
\includegraphics[width=8cm]{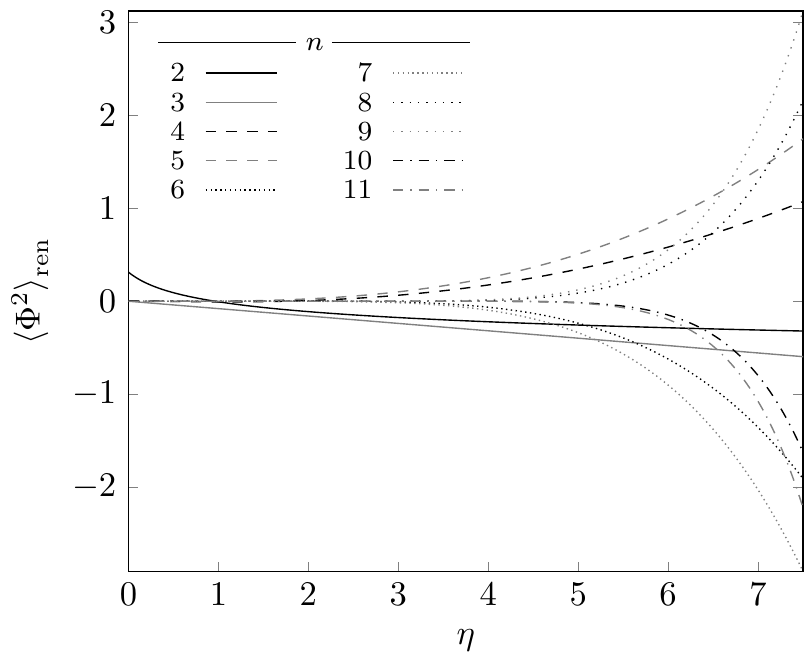}
\end{center}
\caption{$\langle\Phi^{2}_{}\rangle^{}_{\text{ren}}$ for varying $n$ with $a=1$ and $M=\frac{e^{\emc}_{}}{\sqrt{2}}.$\lblf{p}}
\end{figure}
In Minkowski space, the expectation value of the RQFF becomes negligible for large scalar field mass $m$.
In anti-de Sitter space, for fixed curvature coupling $\xi$ we see that the expectation value of the RQFF becomes large in magnitude when the scalar field mass $m$ increases.
This can be understood as follows. The scalar field equation \eqr{sfwe} and the RQFF expectation values \eqr{psr} depend only on $\eta $ and not on the scalar field mass $m$ and curvature coupling $\xi $ separately.
Therefore having large scalar field mass $m$ is equivalent (via \eqr{eta2}) to having small scalar field mass and large negative values of $\xi $.
Therefore, a scalar field having large mass behaves in the same way as a scalar field with a large negative coupling to the negative space-time curvature.

\subsubsection{Varying renormalization mass scale}
For $n$ even, the influence of the renormalization mass scale $M$ on the RQFF can be seen in Fig.~\ref{f:pM}, where we show the RQFF for $M=2^{-\frac{1}{2}}_{}e^{\emc}_{}10^{j-6}_{},$ (with $j=-5,-4,\ldots,5$).
For each even value of $n$, the dependence of the RQFF on the renormalization mass scale is through the $\ln M$ term in $\mathit{\widetilde{\Upsilon\hskip1.2pt}}$ \eqr{upstilde}.
For our choice of values of $M$, the different curves in Fig.~\ref{f:pM} correspond to adding a constant to ${\mathit{\widetilde{\Upsilon\hskip1.2pt}}}$.
For $n=2$, since the RQFF is simply proportional to ${\mathit{\widetilde{\Upsilon\hskip1.2pt}}}$, we therefore find a parallel arrangement of curves.
For even $n>2$, the plots show that the RQFF rapidly increases in magnitude as $\eta$ increases, following the passage of each curve through its zeroes for small values of $\eta$.  For increasing values of $n$ this oscillating phase takes place nearer to $\psr=0$, and the phase where the RQFF has increasing magnitude is delayed and more rapid.  For $n=2p+2$, with $p=1,2,\ldots$, the $j=0$ curve will tend to $(-1)^{p+1}\infty$ as $\eta$ increases.  Consequently, the curves corresponding to $j\gtrless0$ will tend to $\pm(-1)^{p}\infty$.

\begin{widetext}
\begin{figure*}[!]
\begin{center}
\includegraphics[width=8cm]{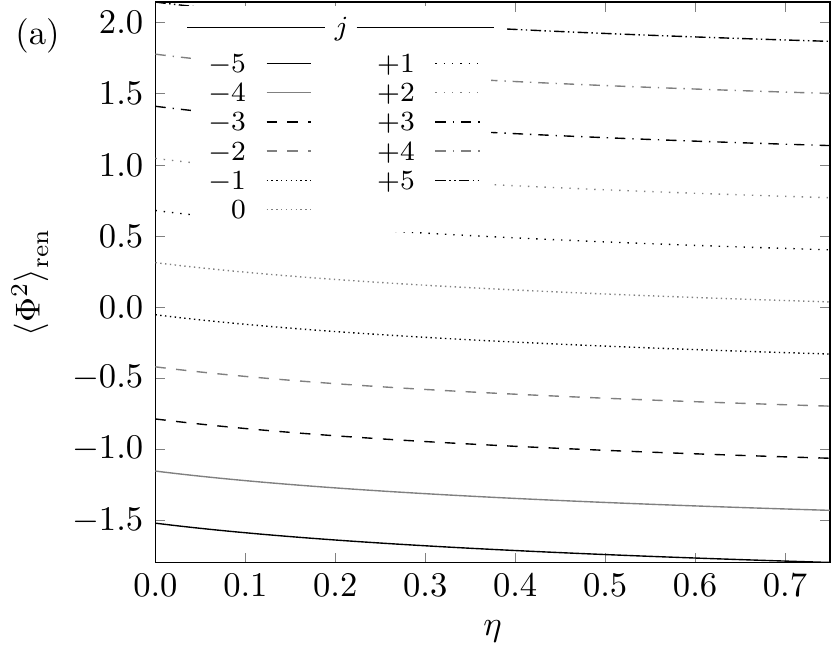}\qquad
\includegraphics[width=8cm]{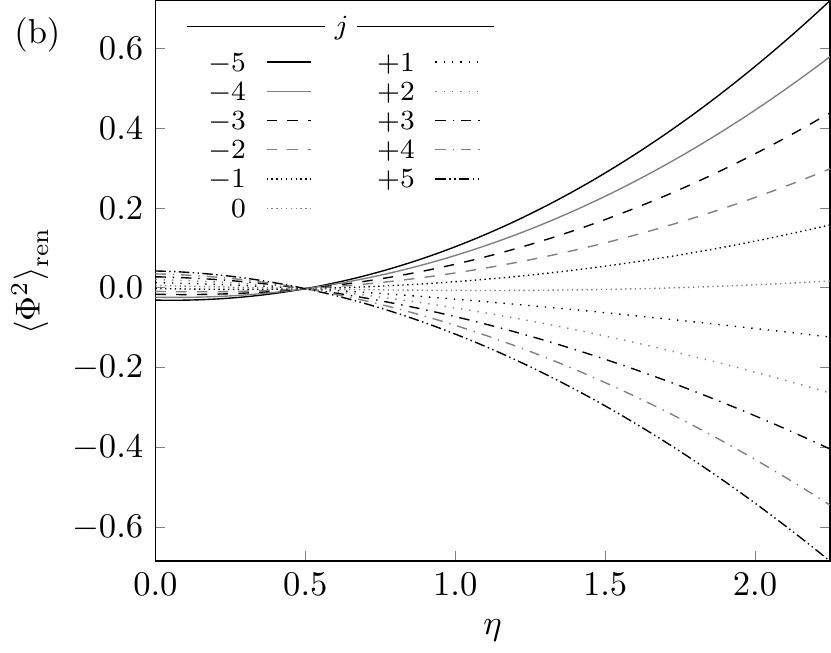}
\end{center}
\begin{center}
\includegraphics[width=8cm]{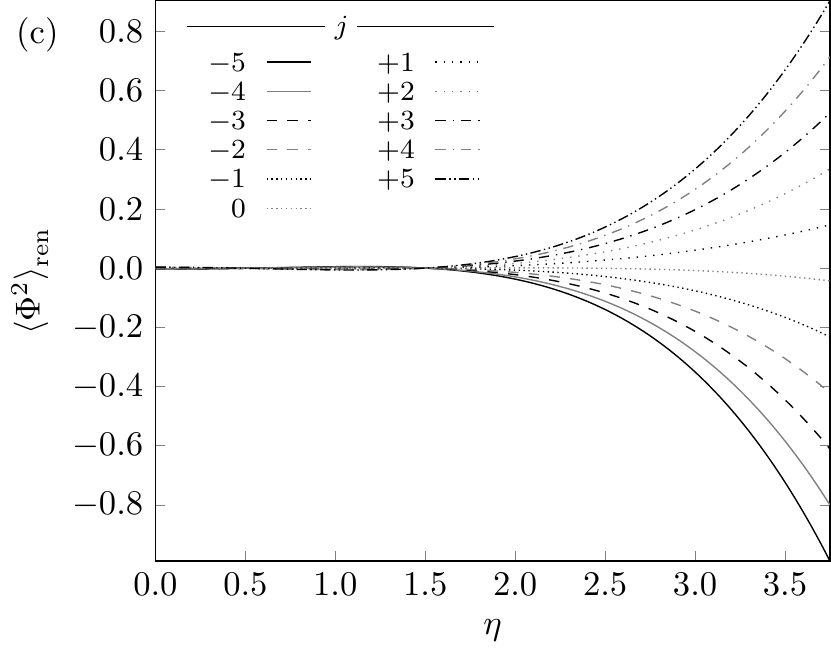}\qquad
\includegraphics[width=8cm]{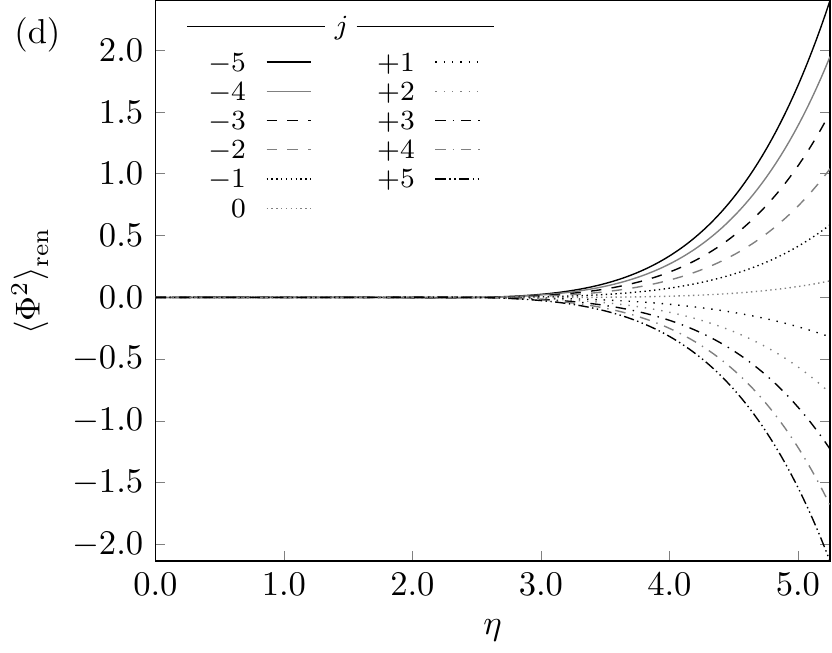}
\end{center}
\begin{center}
\includegraphics[width=8cm]{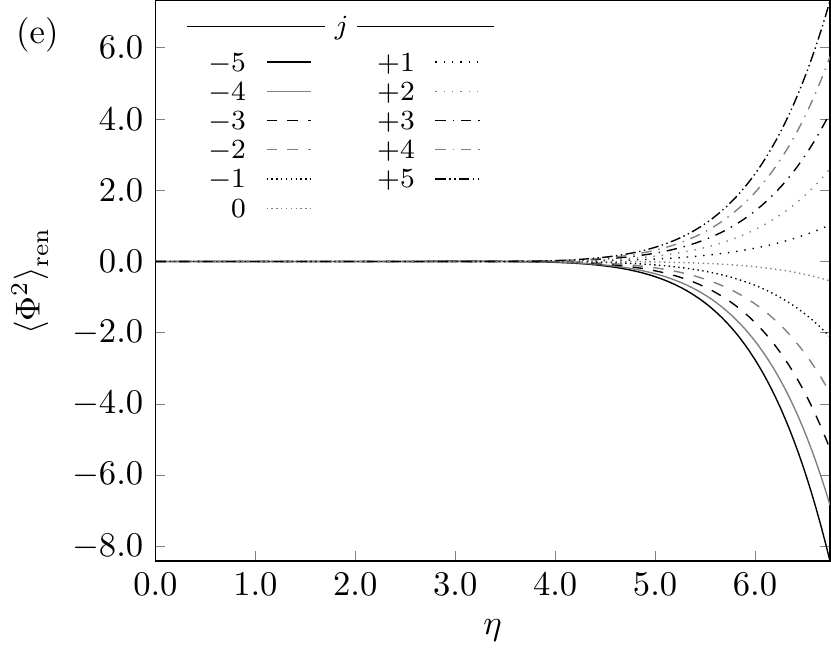}
\end{center}
\bigskip
\caption{$\langle\Phi^{2}_{}\rangle^{}_{\text{ren}}$ for (a) $n=2$, (b) $n=4$, (c) $n=6$, (d) $n=8$ and (e) $n=10$. In each case we have set $a=1$ and $M=\tfrac{e^{\emc}_{}}{\sqrt{2}}10^{j-6}_{}$.\lblf{pM}}
\end{figure*}

\subsection{Results for $\Tmnr$\label{resrset}}
Using (\ref{e:rsetads}, \ref{e:VII}, \ref{e:WII}), expressions for $\Tmnr$ are given below for $n=2$ to $n=11$ respectively:
\begin{subequations}
\label{e:tmn}
\begin{align}
\lble{tmn2}
\Tmnrn^{n=2}=\;&-\frac{1}{8\pi a^{2}}\left[\left(-2\eta^{2}_{}-4\xi+\frac{1}{2}\right)\mathit{\widetilde{\Upsilon\hskip1.2pt}}+\eta^{2}+\frac{1}{12}\right]g^{}_{\mu\nu}+\Theta^{n=2}_{\mu\nu},
\\
\lble{tmn3}
\Tmnrn^{n=3}=\;&\frac{1}{12\pi a^{3}}\left[\eta^{3}+\left(6\xi-1\right)\eta\right]g^{}_{\mu\nu},
\\
\lble{tmn4}
\Tmnrn^{n=4}=\;&\frac{3}{128\pi^{2}a^{4}}\left\{\left[-\frac{4}{3}\eta^{4}-\left(16\xi-\frac{10}{3}\right)\eta^{2}+4\xi-\frac{3}{4}\right]\mathit{\widetilde{\Upsilon\hskip1.2pt}}+\eta^{4}+\left(8\xi-\frac{29}{18}\right)\eta^{2}+\frac{2}{3}\xi-\frac{107}{720}\right\}g^{}_{\mu\nu}+\Theta^{n=4}_{\mu\nu},
\\
\lble{tmn5}
\Tmnrn^{n=5}=\;&-\frac{1}{120\pi^{2}a^{5}}\left[\eta^{5}+\left(20\xi-5\right)\eta^{3}-\left(20\xi-4\right)\eta\right]g^{}_{\mu\nu},
\\
\lble{tmn6}
\Tmnrn^{n=6}=\;&-\frac{11}{4608\pi^{3}a^{6}}\left\{\left[-\frac{12}{11}\eta^{6}-\left(\frac{360}{11}\xi-\frac{105}{11}\right)\eta^{4}+\left(\frac{900}{11}\xi-\frac{777}{44}\right)\eta^{2}-\frac{405}{22}\xi+\frac{675}{176}\right]\mathit{\widetilde{\Upsilon\hskip1.2pt}}\right.\nno\\
&\qquad+\left.\eta^{6}+\left(\frac{270}{11}\xi-\frac{313}{44}\right)\eta^{4}-\left(\frac{435}{11}\xi-\frac{7413}{880}\right)\eta^{2}-\frac{321}{88}\xi+\frac{11969}{14784}\right\}g^{}_{\mu\nu}+\Theta^{n=6}_{\mu\nu},
\\
\lble{tmn7}
\Tmnrn^{n=7}=\;&\frac{1}{1680\pi^{3}a^{7}}\left[\eta^{7}+\left(42\xi-14\right)\eta^{5}-\left(210\xi-49\right)\eta^{3}+\left(168\xi-36\right)\eta\right]g^{}_{\mu\nu},
\\
\lble{tmn8}
\Tmnrn^{n=8}=\;&\frac{25}{147456\pi^{4}a^{8}}\left\{\left[-\frac{24}{25}\eta^{8}-\left(\frac{1344}{25}\xi-\frac{504}{25}\right)\eta^{6}
+\left(\frac{2352}{5}\xi-\frac{2961}{25}\right)\eta^{4}-\left(\frac{21756}{25}\xi-\frac{9687}{50}\right)\eta^{2}
\right.\right. \nno \\
&\qquad\left. +189\xi-\frac{1323}{32}\right] \mathit{\widetilde{\Upsilon\hskip1.2pt}}
+ \eta^{8}+\left(\frac{1232}{25}\xi-\frac{461}{25}\right)\eta^{6}-\left(\frac{8764}{25}\xi-\frac{87983}{1000}\right)\eta^{4}
+\left(\frac{51891}{125}\xi-\frac{3854941}{42000}\right)\eta^{2}
\nno \\
&\qquad\left. +\frac{11969}{300}\xi-\frac{288563}{32000} \right\} g^{}_{\mu\nu}+\Theta^{n=8}_{\mu\nu},
\\
\lble{tmn9}
\Tmnrn^{n=9}=\;&-\frac{1}{30240\pi^{4}a^{9}}\left[\eta^{9}+\left(72\xi-30\right)\eta^{7}-\left(1008\xi-273\right)\eta^{5}+\left(3528\xi-820\right)\eta^{3}-\left(2592\xi-576\right)\eta\right]g^{}_{\mu\nu},
\\
\lble{tmn10}
\Tmnrn^{n=10}=\;&-\frac{137}{14745600\pi^{5}a^{10}}
\left\{\left[-\frac{120}{137}\eta^{10}-\left(\frac{10800}{137}\xi-\frac{4950}{137}\right)\eta^{8}+\left(\frac{226800}{137}\xi-\frac{65835}{137}\right)\eta^{6}
\right. \right. \nno \\
&\qquad\left. -\left(\frac{1332450}{137}\xi-\frac{1296075}{548}\right)\eta^{4}
+\left(\frac{2179575}{137}\xi-\frac{15858315}{4384}\right)\eta^{2}-\frac{7441875}{2192}\xi+\frac{13395375}{17536}\right]\mathit{\widetilde{\Upsilon\hskip1.2pt}} \nno\\
&\qquad+\left.\eta^{10}+\left(\frac{11250}{137}\xi-\frac{20605}{548}\right)\eta^{8}-\left(\frac{207450}{137}\xi-\frac{481133}{1096}\right)\eta^{6}+\left(\frac{3959235}{548}\xi-\frac{161437505}{92064}\right)\eta^{4}\right.\nno\\
&\qquad-\left.\left(\frac{57824115}{7672}\xi-\frac{418944187}{245504}\right)\eta^{2}-\frac{12985335}{17536}\xi+\frac{262292845}{1543168}\right\}g^{}_{\mu\nu}+\Theta^{n=10}_{\mu\nu},
\\
\lble{tmn11}
\Tmnrn^{n=11}=\;&\frac{1}{665280\pi^{5}a^{11}}\big[\eta^{11}+\left(110\xi-55\right)\eta^{9}-\left(3300\xi-1023\right)\eta^{7}+\mathrlap{\phantom{\frac{1}{1}}}(30030\xi-7645)\eta^{5}\nno \\
&\qquad-\left(90200\xi-21076\right)\eta^{3}+\left(63360\xi-14400\right)\eta\big]g^{}_{\mu\nu}.
\end{align}
\end{subequations}
\newpage
For even $n$, the RSET depends on the geometric tensor $\Theta^{}_{\mu\nu}$.
This is evaluated using \eqr{theads}, together with \eqr{VII} and \eqr{WII}, and the resulting
expressions are given below for $n=2,4,6,8$ and $10$:
\begin{subequations}
\begin{align}
\lble{thmn2}
\Theta^{n=2}_{\mu\nu}=&\;-\frac{\ln M^{2}}{8\pi a^{2}}\left[\eta^{2}+2\xi-\frac{1}{4}\right]g^{}_{\mu\nu},
\\
\lble{thmn4}
\Theta^{n=4}_{\mu\nu}=&\;\frac{\ln M^{2}}{64\pi^{2}a^{4}}\left[\eta^{4}+\left(12\xi-\frac{5}{2}\right)\eta^{2}-3\xi+\frac{9}{16}\right]g^{}_{\mu\nu},
\\
\lble{thmn6}
\Theta^{n=6}_{\mu\nu}=&\;-\frac{\ln M^{2}}{768\pi^{3}a^{6}}\left[\eta^{6}+\left(30\xi-\frac{35}{4}\right)\eta^{4}-\left(75\xi-\frac{259}{16}\right)\eta^{2}+\frac{135}{8}\xi-\frac{225}{64}\right]g^{}_{\mu\nu},
\\
\lble{thmn8}
\Theta^{n=8}_{\mu\nu}=&\;\frac{\ln M^{2}}{12288\pi^{4}a^{8}}\left[\eta^{8}+\left(56\xi-21\right)\eta^{6}-\left(490\xi-\frac{987}{8}\right)\eta^{4}+\left(\frac{1813}{2}\xi-\frac{3229}{16}\right)\eta^{2}-\frac{1575}{8}\xi+\frac{11025}{256}\right]g^{}_{\mu\nu},
\\
\lble{thmn10}
\Theta^{n=10}_{\mu\nu}=&\;-\frac{\ln M^{2}}{245760\pi^{5}a^{10}}\left[\eta^{10}+\left(90\xi-\frac{165}{4}\right)\eta^{8}-\left(1890\xi-\frac{4389}{8}\right)\eta^{6}+\left(\frac{44415}{4}\xi-\frac{86405}{32}\right)\eta^{4}
\right.\nno \\
& \qquad -\left.\left(\frac{145305}{8}\xi-\frac{1057221}{256}\right)\eta^{2}+\frac{496125}{128}\xi-\frac{893025}{1024}\right]g^{}_{\mu\nu}.
\end{align}
\end{subequations}
\end{widetext}
\subsubsection{General remarks}
As is the case for the RQFF, for each number of space-time dimension $n$, the expressions (\ref{e:tmn2}--\ref{e:thmn10}) depend on: the length scale $a$ (which determines the $adS^{}_{n}$ radius of curvature); the mass of the field quanta $m$; and $\xi $, the coupling to the space-time curvature. For even $n$, there is an additional dependence on the renormalization mass scale $M$.
As well as depending on $m$ and $\xi $ through the quantity $\eta $ \eqr{eta2}, the expectation values of the
the RSET, unlike those of the RQFF, have an \emph{additional explicit linear dependence on $\xi$}.
Since the $adS^{}_{n}$ vacuum is maximally symmetric, once these parameters are fixed, the RSET equals a constant times the metric tensor $g_{\mu\nu }$.

The results \eqr{tmn} also agree with expressions previously computed by Caldarelli \cite{Cal} using zeta-function regularization methods, up to the addition of a constant, which can be absorbed into the definition of the renormalization mass scale by \eqr{Mren}.

Since the RSET is proportional to the metric tensor, in our plots we show $\tfrac{1}{n}\langle {T_{\mu}^{}}^{\mu}_{}\rangle^{}_{\text{ren}} $, which corresponds to the constant of proportionality. We first plot, in Fig.~\ref{f:t}, the results \eqr{tmn} as functions of $\eta$ for $\xi =0$, with $a=1$ and $M=2^{-\frac{1}{2}}_{}e^{\emc}_{}$.  We recall that the choice of $\xi$ here is for illustrative purposes only.  Since $\xi$ is now fixed however, the quantity $\eta$ is in fact a measure of the field mass $m$.
When $\eta=0$, the RSET vanishes for $n$ odd, but for even $n$, the presence of finite terms means that $\langle{T_{\mu}^{}}^{\mu}_{}\rangle^{}_{\text{ren}}\neq0$.

\begin{figure}[H]
\begin{center}
\includegraphics[width=8cm]{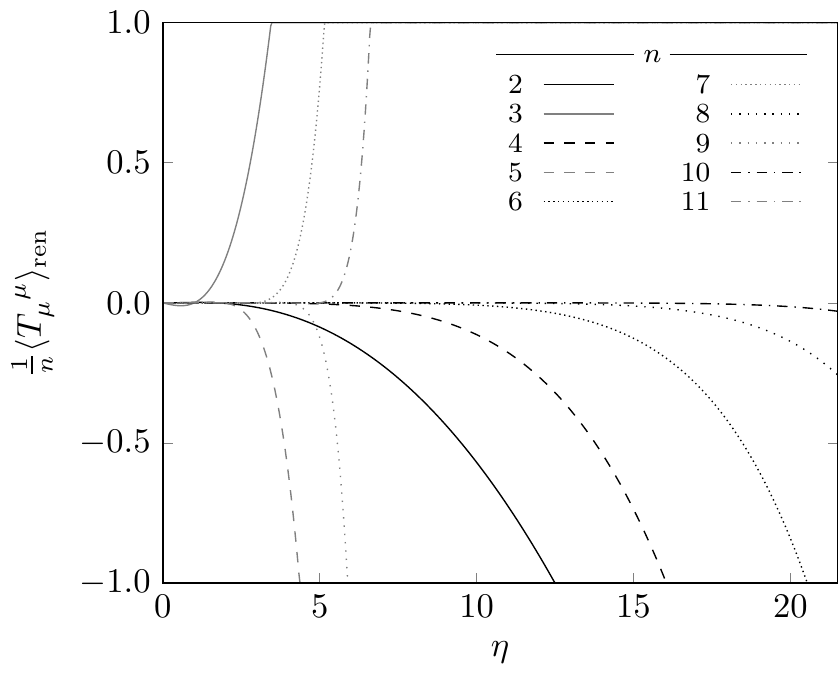}
\end{center}
\caption{$\tfrac{1}{n}\langle {T_{\mu}^{}}^{\mu}_{}\rangle^{}_{\text{ren}}$ with $\xi=0$, $a=1$ and $M=\frac{e^{\emc}_{}}{\sqrt{2}}.$\lblf{t}}
\end{figure}
The curves for even $n$ decrease with increasing $\eta$, whereas those for odd $n$ alternate, being ultimately increasing functions of $\eta $ for $n=3$, $7$ and $11$ and decreasing functions of $\eta $ for $n=5$ and $9$.
As with the RQFF, this behaviour arises from an overall factor of $(-1)^{p}$ in the Green's function for space-times with $2p+1$ and $2p+2$ dimensions which then multiplies the renormalized expectation values (\ref{e:tmn2}--\ref{e:thmn10}). This overall factor can be seen explicitly in the formal Laurent series \eqr{Se} and \eqr{So} which form part of the Green's function.
In addition, curves for even $n$ have a slower rate of change with respect to $\eta$ and can be found on the right-hand side of the plot.  From the expressions for even $n$, it can be understood that their decrease is slowed by the $\psf{\tfrac{1}{2}+\eta}$-term present in $\mathit{\widetilde{\Upsilon\hskip1.2pt}}$.  For increasing $\eta$, after initially passing through its zeroes, the rapid growth of each curve takes place in ascending order of $n$.
The fact that the expectation values $\langle{T_{\mu}^{}}^{\mu}_{}\rangle^{}_{\text{ren}}$ have large magnitude for large $\eta $ with $\xi $ fixed follows from the argument at the end of Sec.~\ref{sec:phi2gen}.
Since the scalar field theory described by \eqr{sfwe} depends only on $\eta $ and not
on the scalar field mass $m$ and curvature coupling $\xi $ separately, a scalar field having large mass $m$ and small coupling $\xi $ behaves in the same way as a field having a small mass and a large negative value of $\xi $.

\subsubsection{Varying renormalization mass scale}
\label{sec:tvarm}
To investigate the effect of changing the renormalization mass scale $M$, in Fig.~\ref{f:tM4} we show
$\frac{1}{4}\langle {T_{\mu}^{}}^{\mu}_{}\rangle^{}_{\text{ren}}$ as a function of $\eta$ for a massive, conformally-coupled field for $n=4$, $a=1$ and with varying $M$.  Once again, recall that this choice is purely for illustration and means that $\eta$ now depends solely on $m$ \eqr{etac}.  We do not include plots for $n$ even with $n\neq4$ since this pattern of behaviour is qualitatively identical to the description given for the plots of $\langle \Phi^{2}_{}\rangle^{}_{\text{ren}}$ in Fig.~\ref{f:pM}.
\begin{figure}[H]
\begin{center}
\includegraphics[width=8cm]{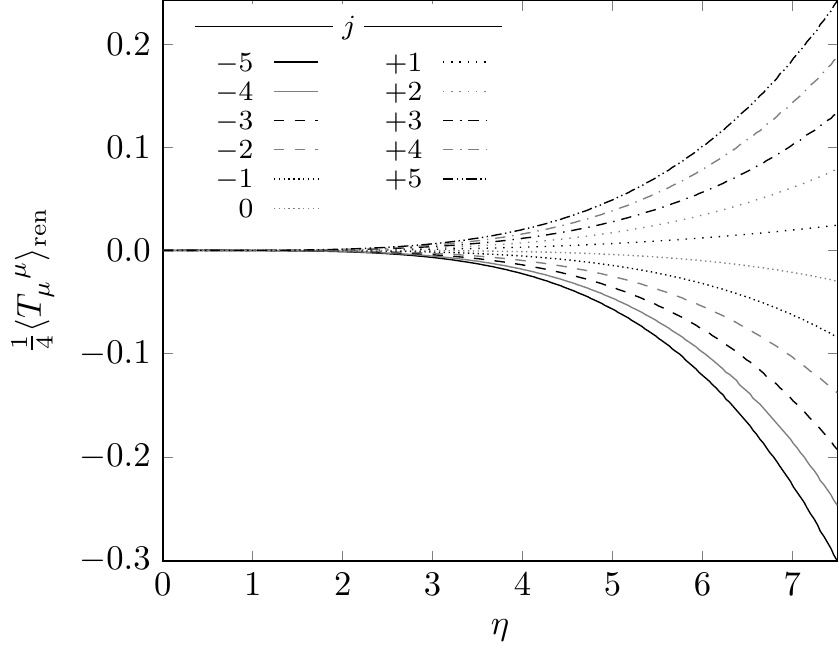}
\end{center}
\caption{$\tfrac{1}{4}\langle {T_{\mu}^{}}^{\mu}_{}\rangle^{}_{\text{ren}}$ for $n=4$, $\xi=\tfrac{1}{6}$, $a=1$, $M=\tfrac{e^{\emc}_{}}{\sqrt{2}}10^{j-6}_{}$.\lblf{tM4}}
\end{figure}

Fig.~\ref{f:tM4} shows that, for fixed curvature coupling $\xi $, the large $\eta$ (or, equivalently, large scalar field mass) behaviour of $\frac{1}{4}\langle {T_{\mu}^{}}^{\mu}_{}\rangle^{}_{\text{ren}}$ (in particular, its sign) depends on the choice of renormalization mass scale. When $n=4$, we see from Fig.~\ref{f:tM4} that small renormalization mass scales give rise to values of $\frac{1}{4}\langle {T_{\mu}^{}}^{\mu}_{}\rangle^{}_{\text{ren}}$ which are negative for large $\eta $ while
large renormalization mass scales give rise to values of $\frac{1}{4}\langle {T_{\mu}^{}}^{\mu}_{}\rangle^{}_{\text{ren}}$ which are positive for large $\eta $.
This behaviour will be the same for $n=8$.
However, for $n=6$ and $n=10$, the opposite behaviour is found: small renormalization mass scales give $\frac{1}{4}\langle {T_{\mu}^{}}^{\mu}_{}\rangle^{}_{\text{ren}}$ positive for large $\eta $ and large renormalization mass scales give $\frac{1}{4}\langle {T_{\mu}^{}}^{\mu}_{}\rangle^{}_{\text{ren}}$ negative for large $\eta $.
This mirrors the behaviour of the RQFF seen in Fig.~\ref{f:pM}.
In Figs.~\ref{f:t} and \ref{f:tmn} we have plotted expectation values for fixed renormalization mass scale.
The results shown in those plots will therefore also change considerably for even $n$ if a different value of the renormalization mass scale is chosen.

\begin{figure*}
\begin{center}
\includegraphics[width=6cm]{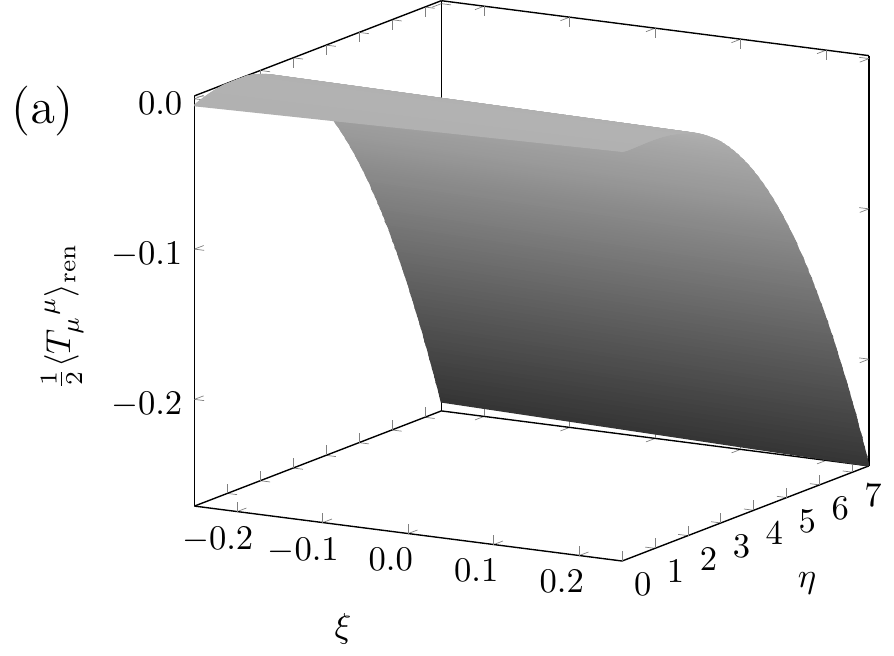}\qquad\quad
\includegraphics[width=6cm]{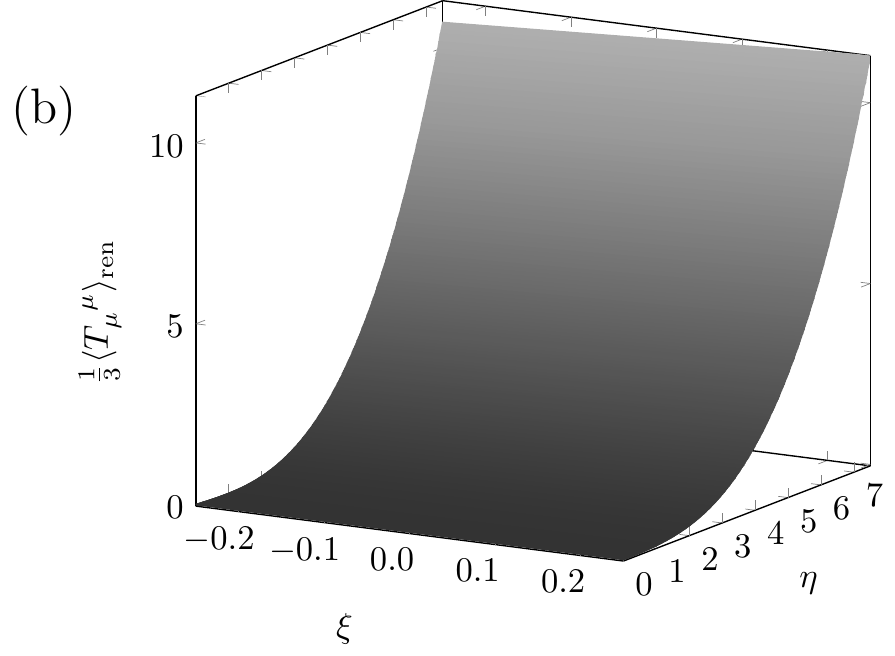}
\includegraphics[width=6cm]{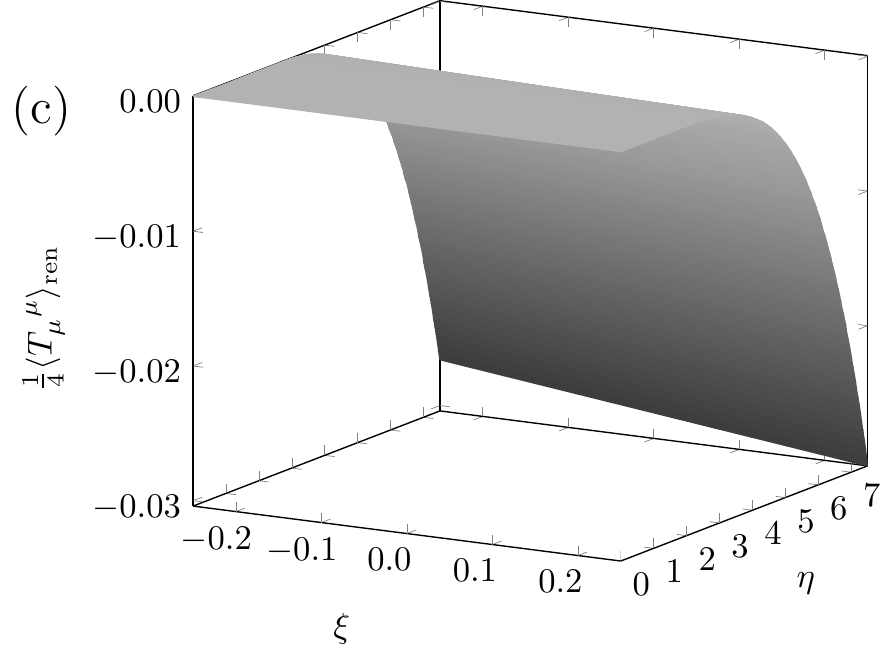}\qquad\quad
\includegraphics[width=6cm]{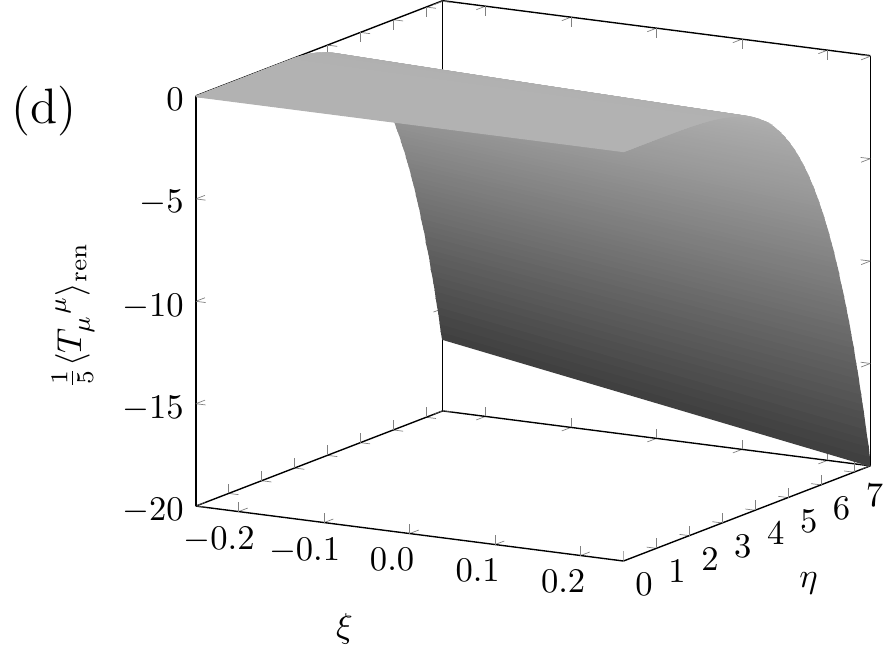}
\includegraphics[width=6cm]{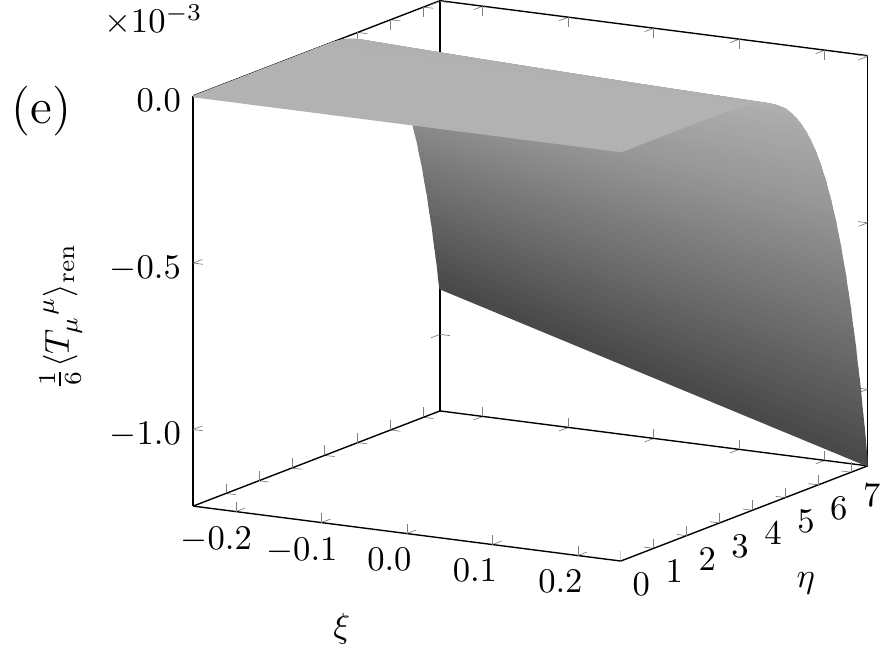}\qquad\quad
\includegraphics[width=6cm]{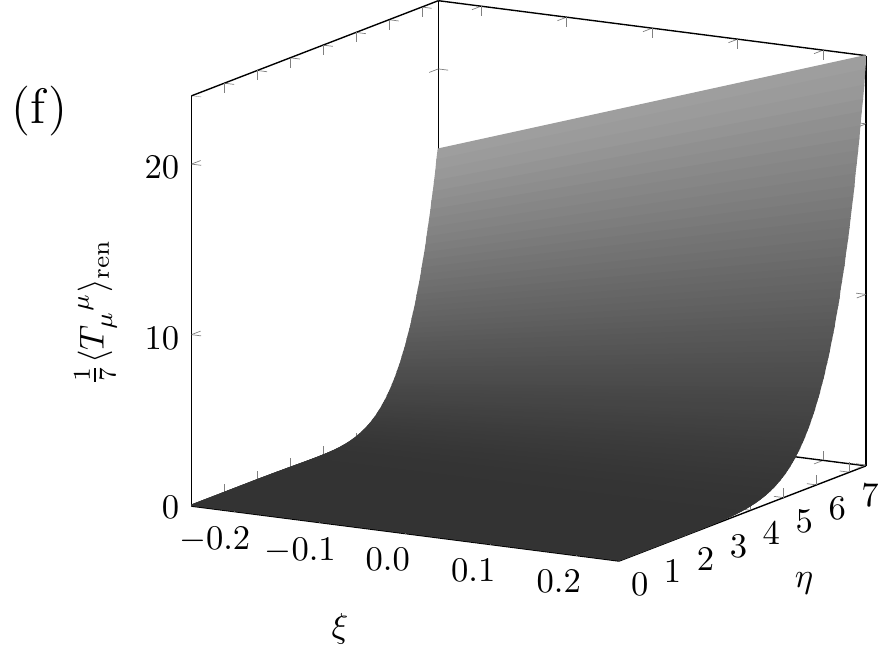}
\includegraphics[width=6cm]{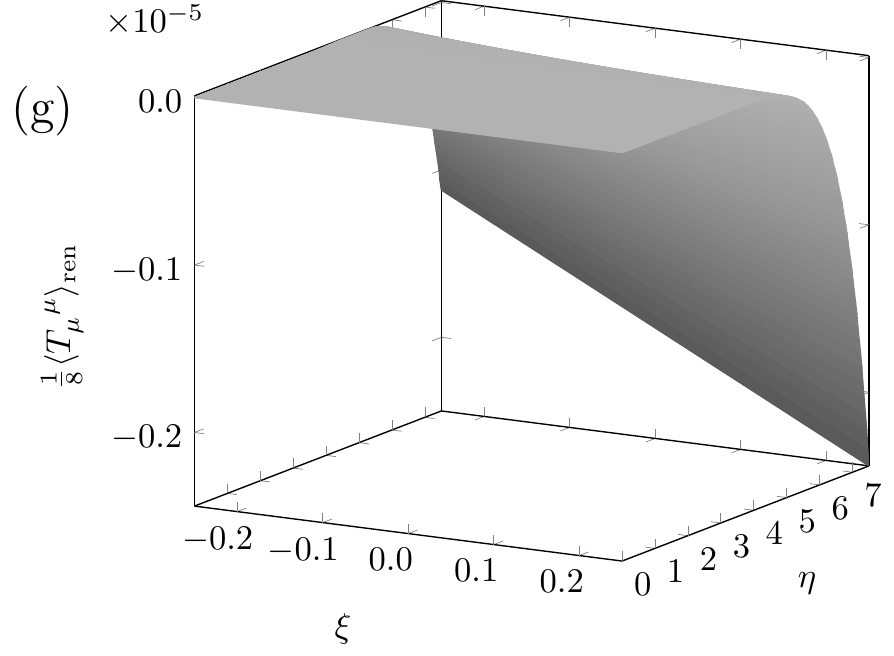}\qquad\quad
\includegraphics[width=6cm]{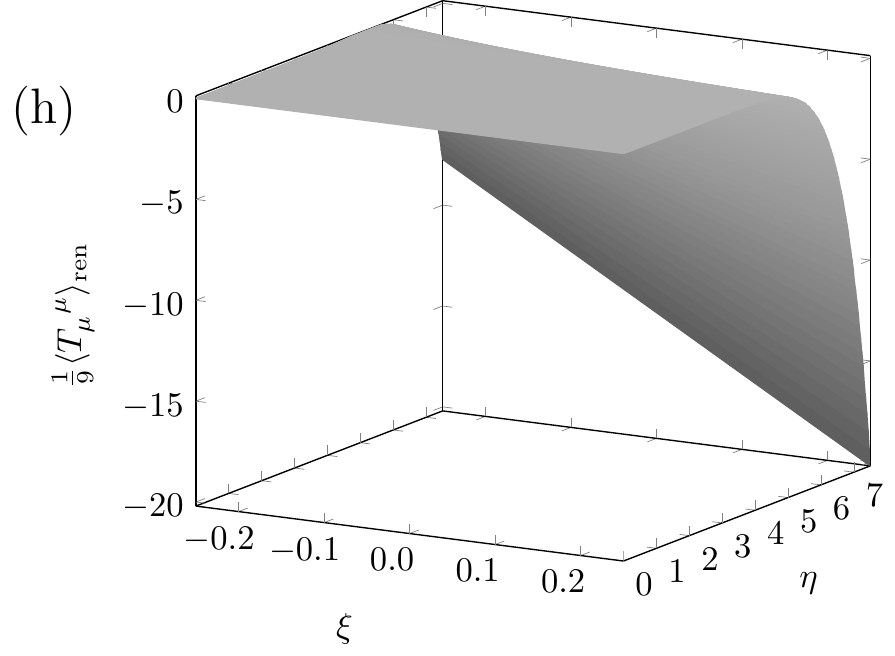}
\includegraphics[width=6cm]{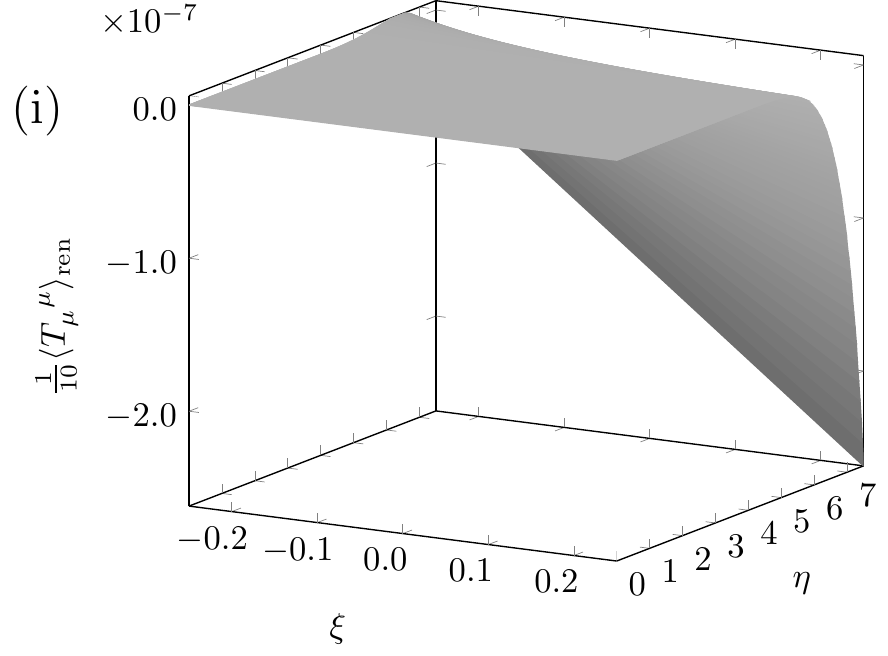}\qquad\quad
\includegraphics[width=6cm]{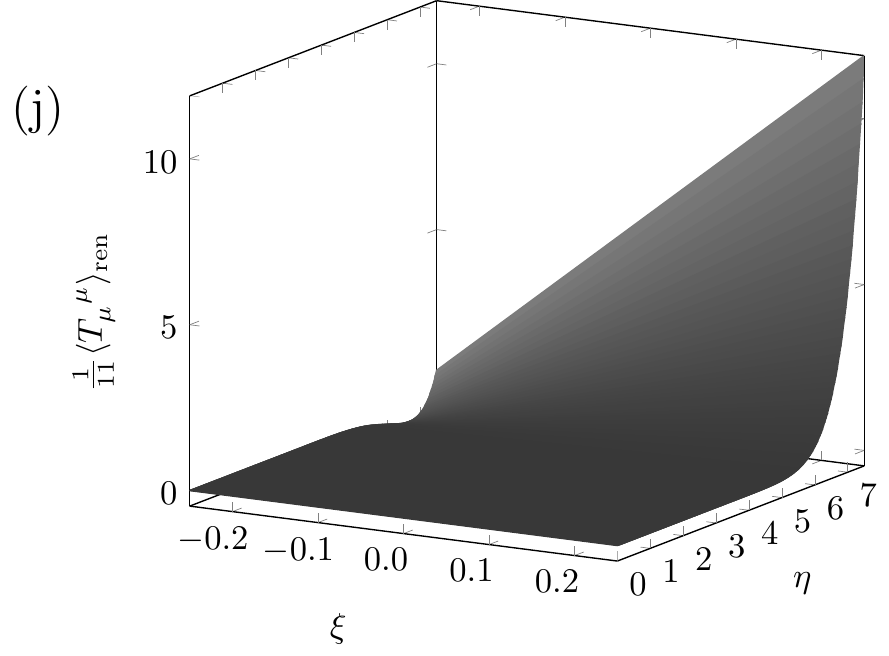}
\end{center}
\caption{$\tfrac{1}{n}\langle{T^{}_{\mu}}^{\mu}_{}\rangle^{}_{\text{ren}}$ as a function of both $\eta$ and $\xi$, with $a=1$
and $M=\frac{e^{\emc}_{}}{\sqrt{2}}$, for (a) $n=2$, (b) $n=3$, (c) $n=4$, (d) $n=5$, (e) $n=6$, (f) $n=7$, (g) $n=8$, (h) $n=9$,
(i) $n=10$ and (j) $n=11$.
\lblf{tmn}}
\end{figure*}

\subsubsection{Varying curvature coupling}
Allowing $\xi $ to vary, in Fig.~\ref{f:tmn} we show $\tfrac{1}{n}\langle {T_{\mu}^{}}^{\mu}_{}\rangle^{}_{\text{ren}}$ as a function of $\eta$ and $\xi$ for $n=2$ to $n=11$ inclusive with $a=1$ and $M=\tfrac{e^{\emc}_{}}{\sqrt{2}}10^{j-6}_{}$.
For $n$ even, the range $|\tfrac{1}{n}\langle {T_{\mu}^{}}^{\mu}_{}\rangle^{}_{\text{ren}}|$ corresponding to $\xi\in[-0.2,0.2]$ with $\eta\in[0.0,7.5]$ decreases as $n$ increases.  Conversely, for $n$ odd, $|\tfrac{1}{n}\langle {T_{\mu}^{}}^{\mu}_{}\rangle^{}_{\text{ren}}|$ increases as $n$ increases.  The linear dependence on $\xi $ for fixed $\eta $ is particularly visible at the intersection of the plot's surface with the $\eta=7.5$ plane.

We remark that the scalar field equation \eqr{sfwe} depends only on the scalar field mass $m$ and curvature coupling $\xi$ through the quantity $\eta $ \eqr{eta2}.
Therefore, at the level of the classical theory, a scalar field with a large mass behaves in the same way as a scalar field with a small mass and large negative coupling to the curvature.
The definition of the RSET \eqr{rsetads} depends explicitly on the curvature coupling $\xi $ as well as $\eta $, yielding the complicated dependence of $|\tfrac{1}{n}\langle {T_{\mu}^{}}^{\mu}_{}\rangle^{}_{\text{ren}}|$ on both $\xi $ and $\eta $ which is depicted in Fig.~\ref{f:tmn}.
The picture is further complicated by the fact that the large $\eta $ behaviour of the expectation values for even $n$ also depends on the choice of mass renormalization scale, as discussed in Sec.~\ref{sec:tvarm}.

\subsubsection{Trace anomaly}
In the particular case that the quantum scalar field is massless and conformally coupled \eqr{mc},
the trace of the geometric tensor $\Theta^{}_{\mu\nu}$ vanishes \cite{D&F08}.
For odd $n$, the vacuum RSET vanishes when the field is massless and conformally coupled.
For even $n$, the trace of the vacuum RSET reduces to the well-known conformal anomaly
(see for example \cite{Duff:1993wm} for a review).
In $adS^{}_{n}$, the explicit expressions for the conformal anomaly for $n=2,4,6,8$ and $10$ are:
\begin{subequations}
\begin{align}
\lble{tmcc2}
\langle{T_{\text{mc }\mu}}^{\mu}\rangle^{n=2}_\text{ren}=\;&-\frac{1}{12\hs a_{}^2\pi},
\\
\lble{tmcc4}
\langle{T_{\text{mc }\mu}}^{\mu}\rangle^{n=4}_\text{ren}=\;&-\frac{1}{240\hs a_{}^4\pi_{}^2},
\\
\lble{tmcc6}
\langle{T_{\text{mc }\mu}}^{\mu}\rangle^{n=6}_\text{ren}=\;&-\frac{5}{4032\hs a_{}^6\pi_{}^3},
\\
\lble{tmcc8}
\langle{T_{\text{mc }\mu}}^{\mu}\rangle^{n=8}_\text{ren}=\;&-\frac{23}{34560\hs a_{}^8\pi_{}^4},
\\
\lble{tmcc10}
\langle{T_{\text{mc }\mu}}^{\mu}\rangle^{n=10}_\text{ren}=\;&-\frac{263}{506880\hs a_{}^{10}\pi_{}^5},
\end{align}
\end{subequations}
where the subscript `mc' indicates the expectation value has been evaluated in the massless, conformally coupled case.
We note that the conformal anomaly is always negative.

\section{Conclusions}
\label{conc}
In this paper we have studied a real massive free quantum scalar field on $adS^{}_{n}$ with arbitrary coupling to the Ricci scalar.
We have used HR to calculate the RQFF $\psr $ and the RSET $\Tmnr$ when the quantum field is in the global $adS^{}_{n}$ vacuum state.
Our method works for arbitrary $n$ and we have presented explicit results for $n=2$ to $n=11$.
The maximal symmetry of both the underlying space-time and the quantum state under consideration have enabled us to give analytic results for all quantities.

We started with a derivation of the $n=4$ Feynman Green's function, $\Gf\xx$ \cite{Campo} that generalizes in a natural way to $n\geq2$.  Due to the maximal symmetry of the space-time and quantum state, $\Gf$  depends only on the distance $s$ between the two space-time points under consideration. We regularized $\Gf$ by expanding it as a formal series to second order in $s$. We have also regularized the Hadamard form $\Gfh$ by expanding it as a formal series to second order in $s$, using the framework in \cite{D&F08}.  We verified that the divergent terms of $\Gf$ match those of $\Gfh$ for $n=2$ to $n=11$ inclusive, and also observed the presence of FRTs for even $n$.
Using these two series expansions, we then computed the RQFF and RSET for each $n$.
Our results for $\psr$ and $\Tmnr$ agree with expressions obtained previously using zeta-function regularization \cite{Cal} up to a choice of renormalization mass scale \cite{Mor}.
The equivalence of zeta-function regularization with HR has been proven rigorously for globally hyperbolic space-times \cite{Mor,Mor1,Hack}.
Even though $adS$ is not globally hyperbolic and so does not satisfy the hypotheses of the theorems in \cite{Mor,Mor1,Hack},
our results have shown that zeta-function regularization and HR remain equivalent on this space-time.

In this paper we have demonstrated how HR works in practice for the vacuum state on $adS^{}_{n}$.
The short-distance divergences characterizing QFTs on curved space-times are independent of the quantum state under consideration.
This means that the process of renormalization for a particular space-time can be performed for a single quantum state, which can be taken to be the simplest, namely the vacuum.
Finding renormalized expectation values for other quantum states then reduces to finding differences in expectation values between two quantum states.
For example, we will consider in a forthcoming publication \cite{KW1+} thermal states on $adS^{}_{n}$.
Such a study will be necessary for future calculations of the RQFF and RSET for a scalar field (with general mass and curvature coupling) on asymptotically $adS^{}_{n}$ black hole space-times.

Finally, let us return to semi-classical Einstein equations \eqr{SEEs}.
Since we have been considering the vacuum state of a free quantum field on a maximally symmetric space-time, our results for $\Tmnr$ have been proportional to the metric tensor
$g_{\mu \nu }$.
The semi-classical Einstein equations \eqr{SEEs} are easily solved exactly in this situation,
simply by making a one-loop quantum correction to the cosmological constant \cite{QFICS}.

\appendix*
\section{}
The calculations that feature in Sec.~\ref{fp} and \ref{ssgf} rely heavily on specific properties of the gamma, psi and hypergeometric functions.  An appropriate formulary is provided here for reference,
throughout which $z,\alpha,\beta,\gamma\in\mathbb{C}$, where $\alpha,\beta,\gamma$ are constants, and $k,j,p\in\{1,2,\ldots\}$.

Firstly, useful results concerning the gamma function, $\gf{z}$ (defined for $z\in\mathbb{C}\setminus\{0,-1,-2,\ldots\}$) are the recurrence relations:
\begin{subequations}
\begin{align}
\lble{grec1}
\gf{z+1}=\;&z\gf{z}=z!,
\\
\lble{grec2}
\gf{z+j}=\;&(z+j-1)(z+j-2)\cdots(z+1)\gf{z+1},
\end{align}
\end{subequations}
and the reflection formula
\be \lble{gref}
\gf{z}\gf{1-z}=\pi\cosec(\pi z),
\ee
(see \S6.1.15 -- \S6.1.17 \cite{A&S}).
Defining the Pochhammer symbol
\be \lble{pdef}
\poch{z}{j}\isdef z\,(z+1)\cdots(z+j-1)=\frac{\gf{z+j}}{\gf{z}},\quad\poch{z}{0}\isdef1,
\ee
(see \S6.1.22 \cite{A&S}), we also have
\be \lble{pd1}
\gf{p+j+z}=\poch{p+z}{j}\poch{z}{p}\gf{z}.
\ee
\quad Useful formulae for the psi functions
\be \lble{psidef}
\psf{z}\isdef\frac{d}{dz}\ln\gf{z},
\ee
(see \S8.361.1 \cite{G&R}) are
\begin{subequations}
\lble{psirec}
\begin{align}
\lble{psirec1}
\psf{j+1}=\;&-\emc+\sum_{l=1}^{j}\frac{1}{l},
\\
\lble{psirec2}
\psf{j+z}=\;&\psf{z}+\sum_{l=0}^{j-1}\frac{1}{z+l},
\end{align}
\end{subequations}
and the reflection result
\be \lble{psiref}
\pi\tan(\pi z)=\psf{\tfrac{1}{2}+z}-\psf{\tfrac{1}{2}-z}
\ee
(see \S8.365.3, \S8.365.4, \S8.365.9 \cite{G&R} respectively), and where in Eq.~\eqr{psirec1} the \emph{boldface} $\emc$ is the Euler-Mascheroni constant.

The hypergeometric \emph{series} (\S15.1.1 \cite{A&S})
\be \lble{hdef}
\hf{\alpha}{\beta}{\gamma}{z}\isdef\sum_{j=0}^{+\infty}\frac{\poch{\alpha}{j}\poch{\beta}{j}}{\poch{\gamma}{j}}\frac{z^{j}_{}}{j!},
\ee
is undefined for $\gamma=0,-1,-2,\ldots$, unless either $\alpha $ or $\beta $ is a negative integer.  The series
converges (diverges) absolutely for $\abs{z}\lessgtr1$. If $z=1$,
\be \lble{AS15.1.20}
\hf{\alpha}{\beta}{\gamma}{z}=\frac{\gf{\gamma}\gf{\gamma-\alpha-\beta}}{\gf{\gamma-\alpha}\gf{\gamma-\beta}},
\ee
for $0<\Real(\gamma-\alpha-\beta) $ (see \S15.1.20 \cite{A&S}).  If $\abs{z}=1$ and $z\neq1$, then $F$
\be
\begin{cases} \lble{AS15.1.1}
\text{converges absolutely},\hfill0<&\hspace{-2.4mm}\Real(\gamma-\alpha-\beta),\\
\text{converges conditionally},\quad\hfill-1<&\hspace{-2.4mm}\Real(\gamma-\alpha-\beta)\leq0,\\
\text{diverges},&\hspace{-2.4mm}\Real(\gamma-\alpha-\beta)\leq-1,
\end{cases}
\ee
(see also \S15.1.1 \cite{A&S}).
Solutions $\hf{\alpha}{\beta}{\gamma}{z}$ of the hypergeometric differential equation (see \S15.5.1 \cite{A&S})
\be \lble{hde}
z(z-1)F^{}_{,\hs zz}+[\gamma-(\alpha+\beta+1)z]F^{}_{,\hs z}-\alpha\beta F=0,
\ee
are the hypergeometric \emph{functions}, exhaustive lists of which can be found in, for example, \cite{A&S, Erd1, G&R} and whose specific forms depend partly on the properties and interrelations of $\alpha$, $\beta$ and $\gamma$.
\begin{widetext}
The linear transformations of hypergeometric series of direct relevance to Sec.~\ref{fp} and \ref{ssgf} are:
\begin{subequations}
\begin{align}
\lble{hlt1}
&\hf{\alpha}{\beta}{\gamma}{z}=(1-z)^{\gamma-\alpha-\beta}_{}\hf{\gamma-\alpha}{\gamma-\beta}{\gamma}{z},\\
\lble{hlt2}
&\hf{\alpha}{\beta}{\gamma}{z}=(1-z)^{-\alpha}_{}\hf{\alpha}{\gamma-\beta}{\gamma}{\tfrac{z}{z-1}};\\
\lble{hlt3}
&\hf{\alpha}{\beta}{\gamma}{z}=\frac{\gf{\gamma}\gf{\gamma-\alpha-\beta}}{\gf{\gamma-\alpha}\gf{\gamma-\beta}}\hf{\alpha}{\beta}{\alpha+\beta-\gamma+1}{1-z}\nno \\
&\mathrlap{\qquad +(1-z)^{\gamma-\alpha-\beta}_{}\frac{\gf{\gamma}\gf{\alpha+\beta-\gamma}}{\gf{\alpha}\gf{\beta}}\hf{\gamma-\alpha}{\gamma-\beta}{\gamma-\alpha-\beta+1}{1-z},}{\hspace{129.25mm}\abs{\arg(1-z)}<\pi,}\\
\lble{CamT1}
&\hf{\alpha}{\beta}{\gamma}{z}=\frac{\gf{\gamma}\gf{\beta-\alpha}}{\gf{\beta}\gf{\gamma-\alpha}}(-z)_{}^{-\alpha}\hf{\alpha}{\alpha+1-\gamma}{\alpha+1-\beta}{z^{-1}_{}}\nno \\
&\mathrlap{\qquad +\frac{\gf{\gamma}\gf{\alpha-\beta}}{\gf{\alpha}\gf{\gamma-\beta}}(-z)_{}^{-\beta}\hf{\beta}{\beta+1-\gamma}{\beta+1-\alpha}{z_{}^{-1}},}{\hspace{132.65mm}\abs{\arg(-z)}<\pi;}\\
\lble{hlt4}
&\hf{\alpha}{\beta}{\alpha + \beta}{z}=\frac{\gf{\alpha+\beta}}{\gf{\alpha}\gf{\beta}}\sum^{+\infty}_{j=0}\frac{\poch{\alpha}{j}\poch{\beta}{j}}{(j!)^{2}_{}}\left[2\psf{j+1}-\psf{\alpha+j}-\psf{\beta+j}-\ln(1-z)\right](1-z)^{j}_{},\nno\\
&\hspace{110mm}\abs{\arg(1-z)}<\pi,\;\abs{1-z}<1,\\
\lble{hlt5}
&\hf{\alpha}{\beta}{\alpha + \beta -j}{z}=(1-z)^{-j}_{}\frac{\gf{j}\gf{\alpha+\beta-j}}{\gf{\alpha}\gf{\beta}}\sum^{j-1}_{k=0}\frac{\poch{\alpha-j}{k}\poch{\beta-j}{k}}{k!\poch{1-j}{k}}(1-z)^{k}_{}-(-1)^{j}_{}\frac{\gf{\alpha+\beta-j}}{\gf{\alpha-j}\gf{\beta-j}}\nno\\
&\qquad \times \sum^{+\infty}_{k=0}\frac{\poch{\alpha}{k}\poch{\beta}{k}}{k!(k+j)!}(1-z)^{k}_{}\left[\ln(1-z)-\psf{k+1}-\psf{k+j+1}+\psf{\alpha+k}+\psf{\beta+k}\right],\nno\\
&\hspace{110mm}\abs{\arg(1-z)}<\pi,\;\abs{1-z}<1,\\
&\hspace{-6.75mm}\text{(see \S15.3.3, \S15.3.4, \S15.3.6, \S15.3.7, \S15.3.10 and \S15.3.12 \cite{A&S}, respectively).}\nno
\end{align}
\end{subequations}
\vspace{1mm}
\begin{acknowledgments}
C.K. would like to thank A. Higuchi and V. Moretti for helpful comments on the theorems concerning the equivalence of zeta-function regularization and HR.

The work of C.K. is supported by EPSRC UK, while that of E.W. is supported by the Lancaster-Manchester-Sheffield Consortium for Fundamental Physics under grant ST/J000418/1 and European Cooperation in Science and Technology (COST) action MP0905 ``Black Holes in a Violent Universe''.
\end{acknowledgments}
\end{widetext}

\pagebreak
%
%
%
\end{document}